\documentclass{aa} 
\usepackage{graphicx}
\graphicspath{{../figures/}}
\usepackage{txfonts}
\usepackage{multirow} 

\begin{document} 

   \title{Study of the optical to X-ray broad emission lines of \object{Mrk\,110}}

   \author{A. Jur\'{a}\v{n}ov\'{a}\inst{1,2},
          E. Costantini\inst{1,2}, 
          L. Di Gesu\inst{3}, 
          J. Ebrero\inst{4}, 
          J. Kaastra\inst{1,5},
          K. Korista\inst{6}, 
          G. A. Kriss\inst{7}, 
          M. Mehdipour\inst{7}, 
          E. Piconcelli\inst{8} \and 
          D. Rogantini\inst{9}
          }

   \institute{SRON Netherlands Institute for Space Research, Niels Bohrweg 4, NL-2333 CA Leiden, the Netherlands\\
              \email{a.juranova@sron.nl}
        \and Anton Pannekoek Institute, University of Amsterdam, Postbus 94249, NL-1090 GE Amsterdam, the Netherlands
        \and Agenzia Spaziale Italiana, Via del Politecnico snc, 00133 Roma, Italy
        \and Telespazio UK for the European Space Agency (ESA), European Space Astronomy Centre (ESAC), Camino Bajo del Castillo, s/n, 28692 Villanueva de la Cañada, Madrid, Spain
        \and Leiden Observatory, Leiden University, PO Box 9513, 2300 RA Leiden, The Netherlands
        \and Department of Physics, Western Michigan University, 1120 Everett Tower, Kalamazoo, MI 49008-5252, USA
        \and Space Telescope Science Institute, 3700 San Martin Drive, Baltimore, MD 21218, USA
        \and INAF--Osservatorio Astronomico di Roma, Via Frascati 33, I-00040 Monte Porzio Catone, Italy
        \and MIT Kavli Institute for Astrophysics and Space Research, Massachusetts Institute of Technology, Cambridge, MA 02139, USA
             }

    \date{Received 26 September 2023; accepted 20 October 2023}
 
  \abstract
   {}
   {In order to shed light on the characteristics of the broad line region (BLR) in a narrow-line Seyfert~1 galaxy, we present an analysis of X-ray, UV, and optical spectroscopic observations of the broad emission lines in Mrk\,110.} 
   {For the broad-band modelling of the emission-line luminosity, we adopt the `locally optimally emitting cloud' approach, which allows us to place constraints on the gas radial and density distribution. By exploring additional environmental effects, we investigate the possible scenarios resulting in the observed spectra.}
   {We find that the photoionised gas in Mrk\,110 responsible for the UV emission can fully account for the observed low-ionisation X-ray lines. The overall ionisation of the gas is lower, and one radial power-law distribution with a high integrated covering fraction $C_{\mathrm{f}} \approx 0.5$ provides an acceptable description of the emission lines spanning from X-rays to the optical band. The BLR is likely more compact than the broad-line Seyfert~1s studied so far, extending from $\sim\!10^{16}$ to $\sim\!10^{18}\,\mathrm{cm}$, and limited by the dust sublimation radius at the outer edge. Despite the large colour excess predicted by the Balmer ratio, the best fit suggests $E(B-V)\approx0.03$ for both the ionising luminosity and the BLR, indicating that extinction might be uniform over a range of viewing angles. While the adopted data-modelling technique does not allow us to place constraints on the geometry of the BLR, we show that the addition of models with a clumpy, equatorial, wind-like structure may lead to a better description of the observed spectra.}
   {}

    \authorrunning{Jur\'{a}\v{n}ov\'{a} et al.}
   \maketitle
   

\section{Introduction}\label{sec:intro}

The broad line region (BLR) is an essential part of the central environment of  active galactic
nuclei \citep[AGNs;][]{Antonucci1993}. The broad emission lines that give this morphological component its name span from the optical band, through UV domain \citep[e.g.][]{Baldwin1975, Peterson1993}, to the X-ray part of the spectrum \citep[e.g.][]{Costantini2007,Peretz2019}, covering a broad range of gas ionisation and density. In Seyfert~1 galaxies, the  full width at half maximum (FWHM) \ion{of the H}{$\beta$} line can reach 10\,000 $\mathrm{km\,s^{-1}}$, and the higher-ionisation lines originating from closer to the central engine may have even larger widths. This distinctive quality is in significant contrast to the line emission from the so-called narrow-line region, where the lines typically have widths of a few hundred $\mathrm{km\,s^{-1}}$ and originate from more distant, lower-density regions, as revealed by the presence of forbidden lines with no broad components \citep{Osterbrock1989}. 

In narrow-line Seyfert~1 galaxies \citep[NLS1s;][]{Goodrich1989}, the BLR emission still appears in the spectra, but the line widths are significantly smaller ($ \mathrm {FWHM_{H \beta}}\lesssim 2000\, \mathrm{km\,s^{-1}} $), which suggests there exists a mechanism that prevents the large broadening from being produced, or at least detected. Relative to the `normal' Seyfert~1s \citep{Jha2022}, NLS1s have been shown to host central black holes of generally lower mass \citep{Wandel1999, Grupe2004a} but higher accretion rate \citep{Sulentic2000}, and show steeper X-ray spectra \citep{Boller1996}. These elements of the primary emission may be connected to the appearance of the broad lines in these objects.

A detailed understanding of the BLR geometry and structure is still elusive, despite substantial efforts from both observational and theoretical perspectives. Nevertheless, significant progress has been made in our understanding of the nature of the BLR \citep[e.g.][]{Peterson1993, Pancoast2014}. It is reasonable to expect the radial extent  of the BLR to be limited by the dusty torus \citep{Netzer1993, Suganuma2006, Landt2014} at the distance where the incident radiation is weak enough to allow the formation and survival of dust particles \citep{Laor1993}. Both line broadening and reverberation mapping studies, which focus on the response of the broad lines to changes in the ionising radiation, point to radial stratification in the gas ionisation. Therefore, unsurprisingly, the lines belonging to more weakly ionised gas show longer delays relative to more highly ionised species. Furthermore, the gas must be located sufficiently close to the nucleus, but must not intercept the line of sight to the source, as the corresponding absorption is absent. Lastly, the line profiles suggest that the dominant motion of the BLR gas is Keplerian, with notable exceptions, where additional components suggest in- or outflowing motion \citep{Laor1993}.

The striking similarity in the broad emission line spectra in the UV and optical bands across different AGN classes led to the proposition of the `locally optimally emitting cloud' (LOC) model \citep{Baldwin1995}. This latter proved to be a simple yet successful approach to describing the line luminosities, without the need for artificial selection effects. The model assumes that the BLR gas spans a wide range of density and ionisation. As a result, it covers optimal conditions for reprocessing of the incoming radiation for each emission line. With this approach, the observed line luminosities can be reproduced if suitable limits for the gas density and distance (ionisation) are used, along with simple weighting of the cloud distribution in both dimensions \citep[e.g.][]{Korista1997a, Goad2012}.

It has been shown that the broad-line region gas can also account for the emission-line fluxes observed in the X-ray band \citep{Costantini2007, Costantini2016}. This concerns mostly the H-like and He-like ions, namely \ion{C}{vi}, \ion{N}{vii}, \ion{O}{viii}, and \ion{O}{vii} and \ion{Ne}{ix}, respectively.

To date, the LOC model has only been successfully applied to the X-ray-to-optical BLR emission of broad-line Seyfert~1 AGN. Extending the global broad-band characterisation to the narrow-line end of Seyfert~1s could provide valuable insights into the general geometrical properties. Therefore, we focus here on Mrk\,110, a nearby \citep[z=0.03529,][]{Keel1996} AGN classified as an NLS1 \citep{Grupe2004}.

The most recent reverberation mapping measurements of the central black hole mass in Mrk\,110 yielded $1.5-3.5 \times 10^7 \,M_{\odot}$ \citep{U2022, Villafana2022}, with a sub-Eddington accretion \citep[$L/L_{\rm Edd}\approx0.4$;][]{Vasudevan2009}. However, we note that the spectropolarimetric analysis of the \ion{H}{$\alpha$} line by \citet{Afanasiev2019} suggests a more massive black hole, namely of $1.3-3.4 \times 10^8 \,M_{\odot}$. The AGN inclination is expected to be relatively small, that is, close to pole-on, with a reported constraint of $37.4^{+9.2}_{-9.5}$ \citep{Wu2001}. The flux in the X-ray \citep{Vincentelli2021}, UV, and optical bands is highly variable on timescales of days to over a decade, and can vary by an order of magnitude in amplitude over longer timescales, both for emission lines and continuum, as recently shown by \citet{Homan2022}.

This paper is structured as follows. In Section \ref{sec:analysis}, we describe our analysis of the X-ray, UV, and optical data used in this study. We focus on the global modelling of the broad lines in Section \ref{sec:fitting} and address possible scenarios that could lead to a deeper understanding of the BLR in this AGN. A discussion of our findings concerning both emission and absorption features is provided in Section \ref{sec:discussion}. Finally, we summarise our conclusions in Section \ref{sec:summary}.
 
Throughout the paper, we adopt the following cosmological parameters: $H_0 = 70 \rm ~km ~s^{-1}~ Mpc^{-1}$, $\Omega_{\rm m} = 0.3$, and $\Omega_{\Lambda} = 0.7$. With these parameters, the luminosity distance to Mrk\,110 is 154~Mpc. For fitting of the X-ray spectra, we used the $C$--statistic \citep{Cash1979, Kaastra2017}, and we used $\chi^2$ for the UV and optical spectral fitting and LOC modelling. The reported uncertainties are calculated at 1$\sigma$ significance. 

\section{Observations and data processing}\label{sec:analysis}
For the purpose of this study, we obtained simultaneous observations with \textit{XMM-Newton} \citep{XMM} and the \textit{Hubble Space Telescope} (HST) Cosmic Origins Spectrograph (COS), allowing us to capture the BLR signatures in the X-ray, UV, and the optical bands. The observation details are listed in Table \ref{tab:COS-XMM}.

\begin{table}[h]
    \centering
    \caption{XMM-Newton (top) and HST-COS (bottom) observation details.
    }
    \begin{tabular}{lccccc}
        \hline\hline

        Observation & Grating/tilt & Date & Exposure \\
                    &              &      & (s)      \\
        \hline
        0840220701 &     $\cdots$      & 2019-Nov-03 & 43\,600 \\
        0840220801 &     $\cdots$      & 2019-Nov-05 & 43\,000 \\
        0840220901 &     $\cdots$      & 2019-Nov-07 & 40\,600 \\
        \hline
        ldye01010 & G130M/1222 & 2019-Nov-03 & 1232 \\
        ldye01020 & G160M/1533 & 2019-Nov-03 & 1248 \\
        ldye01030 & G160M/1589 & 2019-Nov-03 & 1200 \\
        \hline
    \end{tabular}
    \label{tab:COS-XMM}
\end{table}

\subsection{X-ray observations}
The \textit{XMM--Newton} observations were taken during three consecutive revolutions, totalling over 120 ks of data. We processed them following the standard procedures of the \textit{XMM--Newton} Science Analysis System (version 20.0.0), using the most recent calibration files (available on 6 July 2021). For the following analysis of the X-ray emission lines, we used the resulting first-order Reflection Grating Spectrometer \citep[RGS;][]{RGS} spectra, stacking together the data from both RGS 1 and RGS 2 from the three observations.

\begin{figure*}[ht]
    \centering
    \resizebox{\hsize}{!}{\includegraphics{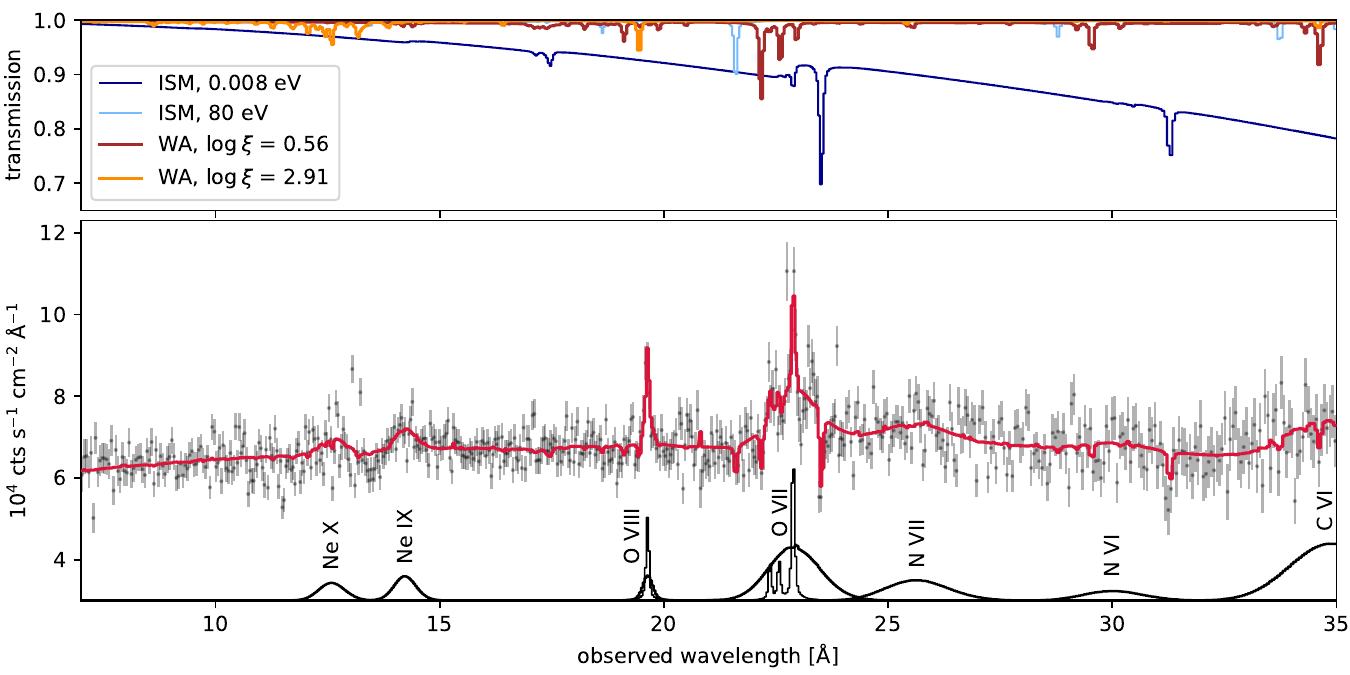}}
    \caption{Fit of the RGS spectrum of Mrk\,110. {\it Bottom panel}: Observed spectrum is overlaid with the best-fitting model ($C$-stat $= 1038$ for $900$ degrees of freedom, red line). The individual emission line components of the model are plotted in black at the bottom of this panel (with a vertical offset of 3 for visualisation purposes). The data were rebinned for clarity. {\it Top panel}: Transmissivity of the absorbing components: Galactic absorption (dark and light blue) and ionised absorbers in Mrk\,110 with $\log \xi = 0.56 $ (brown), $\log \xi = 2.91 $ (orange).}
    \label{fig:RGS}
\end{figure*}

We analysed the X-ray spectrum (Fig. \ref{fig:RGS}) with the fitting package \textsc{SPEX} v3.07.01 \citep{SPEX, SPEX30701}. To model the $7-35$~\AA\ high-resolution RGS spectrum, we composed our model in the following way. For the continuum, a single power law with a slope of $\Gamma = 2.161\pm 0.009$ provided an adequate description. The Galactic absorption is well represented by a \texttt{hot} component with a hydrogen column density of $1.35\times 10^{20}~ \rm cm^{-2}$ \citep{LAB} and a temperature of $kT \sim 8\times 10^{-3}\rm ~eV$, ensuring all atoms are in the neutral state. Additionally, we allowed for another, warmer component ($kT= 80\rm ~eV$) to account for Galactic absorption resulting from more energetic transitions, particularly of \ion{O}{VII}, \ion{N}{VI}, and \ion{C}{VI}. Our best fit yields a column density of the hot interstellar medium of $\sim\! 1.07\times 10^{19}~ \rm cm^{-2}$. 

In addition to the Galactic absorption, we tested for the presence of absorbing photoionised gas in Mrk\,110. For this purpose, we used the \texttt{xabs} model with the ionic column densities derived using the ionising continuum described in Sect. \ref{sec:SED} and \textsc{Cloudy} \citep[v22.00][]{Ferland2017} to determine the ionisation balance. We detected two photoinised components with $\log \xi = 0.6_{-0.2}^{+0.3} $ and $\log \xi = 3.0_{-0.2}^{+0.1} $, respectively, where $\xi$ is the \citet{Kallman2001} definition,

\begin{equation}\label{eq:xi}
    \xi = \frac{L_{\mathrm{ion}}}{nr^2},
\end{equation}

\noindent expressed in units of $\rm erg~s^{-1}~cm $, where $n$ is the hydrogen number density, $L_{\mathrm{ion}}$ is the ionising luminosity in the range 1--1000\,Ryd, and $r$ is the gas distance from the source of the ionising radiation.
The respective line-of-sight outflow velocities are $- 2600 \pm 200$ and $-2900_{-800}^{+400} ~\rm km\,s^{-1}$, and the column densities are $(3\pm 1) \times 10^{19}~\rm cm^{-2}$ and $<5 \times 10^{20}~\rm cm^{-2}$, respectively. The transmissivity of all absorbing components in the RGS spectrum is plotted in the top panel of Fig. \ref{fig:RGS}.

The brightest emission lines are clearly a composition of rather broad and much narrower features. The unresolved narrow emission lines of the \ion{O}{VII} triplet at $\sim\!23$~\AA\ in the observer frame were modelled with delta-line profiles, limiting their widths to the instrumental resolution as a result. Assuming they originate in a tenuous ($<10^9~\rm cm^{-3}$) photoionised gas of the narrow-line region, we fixed the ratio of the resonance and intercombination lines to the forbidden line to 1:4 \citep{Porquet2000}. Similarly, a narrow component was added also for the \ion{O}{VIII} Ly\,$\alpha$ to account for the observed line profile. The remaining emission lines were sufficiently described with a single Gaussian component each, with the line normalisation and width free to vary. For well-constrained lines, the centroid energy was also left free, but it was kept fixed to the theoretical values where only upper limits for the line fluxes were obtained. Additionally, an upper limit of 19\,000 km\,s$^{-1}$ was placed on the FWHM, corresponding to three times the value for the broadest \ion{Ly}{$\alpha$} component, as described below.

The resulting fluxes and line widths are summarised in Table \ref{tab:lines-X-ray} and the line profiles are visualised individually below the observed spectrum in the bottom panel of Fig. \ref{fig:RGS}. Notably, centroid redshifts of $7000\pm1000 ~\rm km\,s^{-1}$ are detected for both \ion{O}{VII} and \ion{Ne}{IX}. The centroid locations of the remaining lines are consistent with rest-frame energies within the uncertainties.

To model the spectral energy distribution (SED, Sect. \ref{sec:SED}) of the ionising radiation, we also  processed the \textit{XMM--Newton} EPIC-pn data from obs. 0840220701, which was taken immediately after the COS measurements, and used the resulting spectrum to constrain the X-ray continuum. The pn observations were obtained in the small window mode with a thick filter, and they are not affected by a detectable pileup. The events recorded during read-out and high-background periods were excluded, the latter based on a count-rate threshold of $0.2~\rm ct\,s^{-1}$, leaving the net exposure time of 42.5~ks. Additionally, the calibrated event lists were further filtered, leaving only single and double events. The background spectra were extracted from the same chip, avoiding contamination from the source and the bulk of out-of-time events.

\begin{table}
\caption{Properties of the X-ray broad emission lines.}
\centering\label{tab:lines-X-ray}
\begin{tabular}{lc|cc}
\hline\hline
Line & $\lambda_0 $ & FWHM & Flux \\
 & \AA & $\rm km\,s^{-1}$ & $\mathrm{10^{-14}\,erg\,s^{-1}\,cm^{-2}}$ \\
\hline
Ne X   & 12.14 & $ 15\,000_{-8000}^{+4000} $ & $ 5\pm 2 $ \\
Ne IX  & 13.45 & $ 11\,000_{-3000}^{+4000} $ & $ 6 \pm 1 $ \\
O VIII & 18.97 & $ 3000_{-1000}^{+8000} $ & $ 2 \pm 1 $ \\
O VII  & 21.60 & $ 17\,000 \pm 3000 $ & $ 17 \pm 3 $ \\
N VII  & 24.78 & $ <19\,000 $ & $ 7 \pm 2 $ \\
N VI   & 29.53 & $ <15\,000 $ & $ 3_{-2}^{+3} $ \\
C VI   & 33.74 & $ <19\,000 $ & $ 19 \pm 3 $ \\
\hline
\end{tabular}
\tablefoot{The flux given in the last column is the line flux corrected for absorption, measured at Earth's distance (154~Mpc).}
\end{table}

\subsection{UV and optical observations}\label{ssec:UVoptdata}
The HST COS observations were taken with the Primary Science Aperture and gratings G130M and G160M, covering the far-ultraviolet (FUV) radiation between 1100 and 1760 \AA. The data were processed with the calibration pipeline CalCOS v3.3.10. We corrected the observed spectrum for Galactic extinction, $E(B-V)=0.011$ \citep{Schlafly2011} in the direction of Mrk\,110, using a Galactic extinction curve \citep{Fitzpatrick1999} with $R_V=3.1$. The effect of additional reddening in the host galaxy is discussed in detail in Sect. \ref{ssec:extinction}.

In addition to the X-ray and FUV analysis, we extended our focus to longer wavelengths to include more of the BLR emission. We processed the \textit{XMM--Newton} Optical Monitor \citep[OM,][]{OM} grism observations to verify that archival HST Space Telescope Imaging Spectrograph (STIS) spectra could be used to analyse the spectra in the optical band. In addition to the higher spectral resolution and larger effective area of STIS, these observations were also favoured because of calibration issues in the UV band of OM, enabling us to place tighter constraints on the emission-line properties. The STIS spectrum used in our analysis is a composite of the most recent  available near-UV--optical observations obtained during December 2017 and January 2018 \citep{Vincentelli2021}.

To properly determine the underlying continuum and isolate the broad emission lines, we used a model based on the template of \citet{Tsuzuki2006} to represent the moderate blended \ion{Fe}{II} emission originating in the BLR. The model, covering the energy range of $2280-5800$~\AA, was convolved with a Gaussian profile to match the broadening of the \ion{Fe}{II} features in the spectrum and its normalisation was determined during the line-fitting process detailed below.

Having accounted for the contaminating signal, we proceeded to the modelling of the UV and optical BLR lines. Depending on the observed profile, each emission line required up to four Gaussian components. For doublets, we modelled each component separately, whenever distinguishable in the spectrum, tying widths of the doublet components together and their relative distance to the theoretical value. Otherwise, and also for other blended lines or multiplets, we modelled the combined emission and treated the resulting flux accordingly further on in the modelling.

We categorise the line components according to their width as narrow, intermediate, broad, and very broad (Table \ref{tab:lines-UVoptical}). The narrow component, with a full width at half maximum (FWHM) of typically $500-800~\rm km~s^{-1}$, assuming broadening due to Keplerian motion, was required in nearly all studied lines. We did not consider this emission further, assuming it to be coming from the narrow-line rather than the broad-line region.

The broad lines in Mrk\,110 are generally rather narrow. Therefore, BLR emission is manifested through components as narrow as $ 1200~\rm km~s^{-1} \lesssim$ FWHM $\lesssim 2200~\rm km~s^{-1}$. This intermediate-width component constitutes a significant fraction of the flux of the strongest lines; in addition, it proved necessary also for the modelling of some less prominent emission features.

The line components with $ 2200~\rm km~s^{-1} \lesssim$ FWHM $\lesssim 7000~\rm km~s^{-1}$, referred to as `broad' in the context of this work, are sufficient to describe the extended wings of most of the broad-line features. However, the strongest hydrogen lines, that is \ion{Ly}{$\alpha$} and \ion{H}{$\alpha$}, and the \ion{C}{IV} doublet, require an additional (`very broad') component, with FWHM $> 7000~\rm km~s^{-1}$. Interestingly, the centroid wavelengths of all very broad components are redshifted. The largest redshift, corresponding to $(5000\pm200)~\rm km~s^{-1}$, is displayed by \ion{C}{iv}, followed by \ion{H}{$\alpha$} with $(1600\pm100)~\rm km~s^{-1}$ and \ion{Ly}{$\alpha$} with $(400\pm80)~\rm km~s^{-1}$. Generally, the broad components of other lines are also centred redwards of their narrow counterparts, but the redshift is smaller, typically $200 - 600~\rm km~s^{-1}$.

\begin{figure*}[ht]
    \centering
    \resizebox{0.75\hsize}{!}{\includegraphics{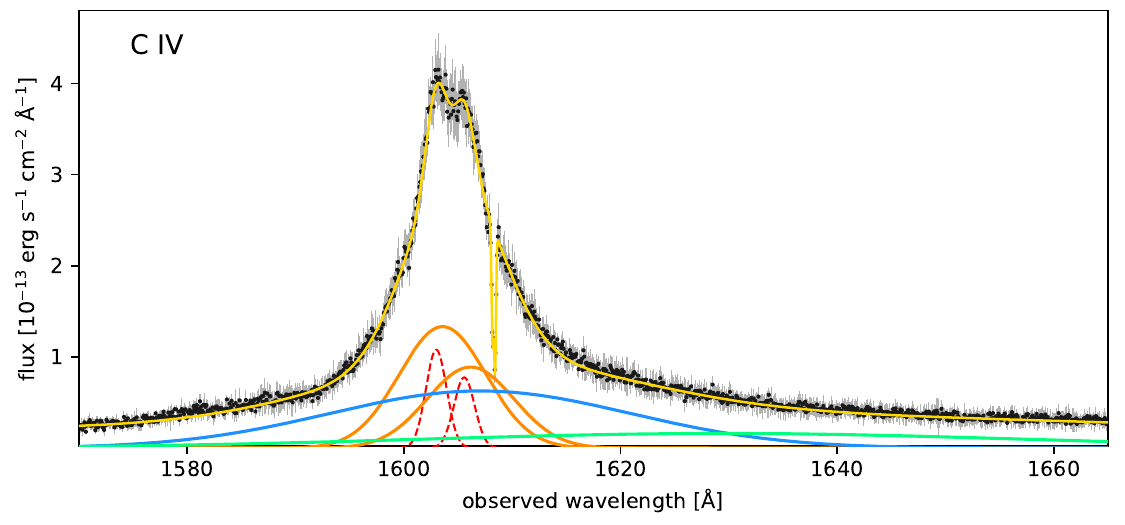}} 
    \caption{\ion{C}{IV} 1548~\AA\ doublet modelled with Gaussian lines. The narrow (dashed red line) and the intermediate (orange) model components were represented for each line of the doublet separately, assuming the same line widths. The remaining wider Gaussians (blue and green) each represent the doublet as a single component. The data were rebinned for clarity.}
    \label{fig:CIV}
\end{figure*}

An example of the line profile fit is given in Fig. \ref{fig:CIV}, where the observed \ion{C}{IV} doublet is plotted together with the fitted line profile and the individual components that form it. The fluxes of the narrow and intermediate doublet components were modelled individually, with their width tied and their relative separation fixed to the theoretical value. However, for the broad and very broad components, a single Gaussian was used for the blue and red component together, as they were inseparable in the spectrum due to their large widths. Finally, the narrow absorption feature in the red wing of the \ion{C}{IV} profile was identified as an interstellar \ion{Fe}{II} line.

The fluxes and widths of the broader components are listed in Table \ref{tab:lines-UVoptical}. We note that in the case of blends or close multiplets, the presented fluxes (and widths in most cases) represent the properties of the entire emission features rather than individual lines contributing to them (see the notes for Table 3). The possibility of weaker nearby lines blending with the listed ones was taken into account in the modelling accordingly.

In cases where the relative statistical uncertainties associated with the line fluxes are only a few percent, the total uncertainty is likely larger \citep{Costantini2016}, given that the flux is derived assuming a well-determined continuum level, nearby lines, and other emission components as described above. Therefore, we artificially increase the uncertainty to 20 \% of the associated flux for the subsequent BLR modelling, which corresponds to the average relative contribution of the narrow components to the fitted total line fluxes.

\begin{table*}[ht]
\caption{Properties of the UV-optical broad emission line components derived from the HST COS (top panel) and STIS (bottom) spectra.}
\centering\label{tab:lines-UVoptical}
\begin{tabular}{lc|cc|cc|cc}
\hline\hline
Component: && \multicolumn{2}{c}{Intermediate} & \multicolumn{2}{c}{Broad} & \multicolumn{2}{c}{Very broad} \\
\hline
Line & $\lambda_0$ [\AA] & FWHM & Flux & FWHM & Flux & FWHM & Flux \\
\hline

Ly $\alpha$ & 1216 & $ 1470 \pm 20$ & $ 258 \pm 4$ & $ 3600 \pm 70$ & $ 250 \pm 7$ & $ 8800 \pm 100$ & $ 200 \pm 10$ \\
\ion{N}{V}\tablefootmark{a} & 1239 & $\cdots$ & $\cdots$ & $ 3530 \pm 70$ & $ 37 \pm 3$ & $\cdots$ & $\cdots$ \\
\ion{Si}{II}\tablefootmark{b} & 1260 & $\cdots$ & $\cdots$ & $ 3800 \pm 800 $ & $ <20 $ & $\cdots$ & $\cdots$ \\
\ion{O}{I} \tablefootmark{b} & 1302 & $\cdots$ & $\cdots$ & $ 2390 \pm 40$ & $ 23.4 \pm 0.8$ & $\cdots$ & $\cdots$ \\
\ion{C}{II}\tablefootmark{b} & 1335 & $ 1800 \pm 400$ & $ 2 \pm 1$ & $\cdots$ & $\cdots$ & $\cdots$ & $\cdots$ \\
\ion{Si}{IV}, \ion{O}{IV]}, \ion{S}{IV]}\tablefootmark{b} & 1394 & $\cdots$ & $\cdots$ & $ 5800 \pm 100$ & $ 37 \pm 1$ & $\cdots$ & $\cdots$ \\
\ion{N}{IV]} & 1486 & $\cdots$ & $\cdots$ & $ 2200 \pm 100$ & $ 4.0 \pm 0.5$ & $\cdots$ & $\cdots$ \\
\ion{C}{IV}\tablefootmark{c} & 1548 & $ 1760 \pm 20$ & $ 220 \pm 40$ & $ 6090 \pm 80$ & $ 218 \pm 4$ & $ 12000 \pm 200$ & $ 112 \pm 5$ \\
\ion{He}{II}, \ion{O}{III]}\tablefootmark{c} & 1640 & $\cdots$ & $\cdots$ & $ 3000 \pm 100$ & $ 52 \pm 2$ & $\cdots$ & $\cdots$ \\\hline
\ion{Si}{III]}, \ion{C}{III]}\tablefootmark{b,e} & 1909 & $\cdots$ & $\cdots$ & $ 5400 \pm 200$ & $ 35 \pm 3$ & $\cdots$ & $\cdots$ \\
\ion{Mg}{II}\tablefootmark{d} & 2796 & $\cdots$ & $\cdots$ & $ 2810 \pm 40$ & $ 58 \pm 2$ & $\cdots$ & $\cdots$ \\
H $\delta$ & 4102 & $\cdots$ & $\cdots$ & $ 3300 \pm 100$ & $ 6.1 \pm 0.5$ & $\cdots$ & $\cdots$ \\
H $\gamma$ & 4340 & $\cdots$ & $\cdots$ & $ 4600 \pm 200$ & $ 8.6 \pm 0.8$ & $\cdots$ & $\cdots$ \\
\ion{He}{II} & 4686 & $\cdots$ & $\cdots$ & $ 5500 \pm 200$ & $ 10.3 \pm 0.5$ & $\cdots$ & $\cdots$ \\
H $\beta$ & 4861 & $ 1710 \pm 40$ & $ 18.9 \pm 0.7$ & $ 4400 \pm 100$ & $ 16 \pm 1$ & $\cdots$ & $\cdots$ \\
\ion{He}{I} & 5876 & $ 2100 \pm 100$ & $ 5.2 \pm 0.3$ & $ 3400 \pm 700$ & $ 2.0 \pm 0.4$ & $\cdots$ & $\cdots$ \\
H $\alpha$ & 6563 & $ 1800 \pm 40$ & $ 98 \pm 4$ & $ 3300 \pm 200$ & $ 42 \pm 8$ & $ 7800 \pm 400$ & $ 29 \pm 3$ \\

\hline
\end{tabular}
\tablefoot{ The FWHM is given in units of $\rm km\,s^{-1}$ and the flux in $\mathrm{10^{-14}\,erg\,s^{-1}\,cm^{-2}}$, measured at Earth's distance (154~Mpc). The flux is corrected for line absorption and Galactic dust extinction.
\tablefoottext{a}{Doublet, components modelled separately.}
\tablefoottext{b}{Blend, modelled with one Gaussian component.}
\tablefoottext{c}{Doublet, intermediate components modelled separately, broad and very broad together as one Gaussian each.}
\tablefoottext{d}{Doublet, broad lines modelled as one component.}
\tablefoottext{e}{Possible contribution from a feature from the Fe III UV34 multiplet, as explained in \citet{Leighly2004}.}
}
\end{table*}

\section{Modelling the broad line region }   
\subsection{Incident continuum spectral energy distribution}\label{sec:SED}

\begin{figure}[ht]
    \centering
    \resizebox{\hsize}{!}{\includegraphics{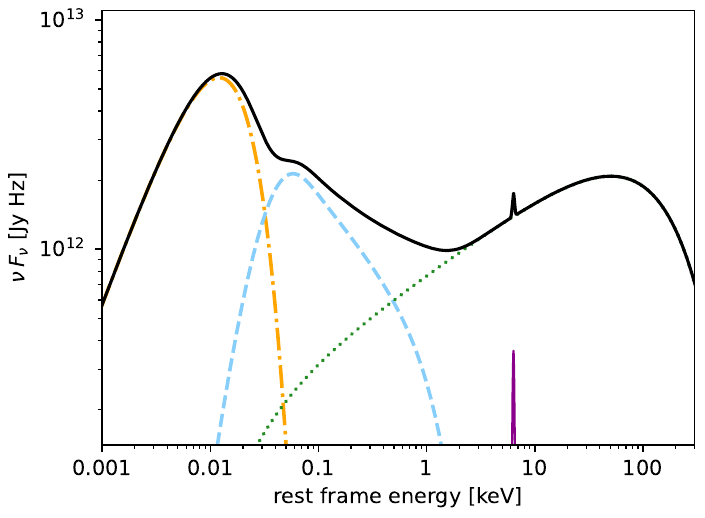}}
    \caption{Incident continuum SED of Mrk\,110 (black line) obtained from a fit of HST continuum data points and the {\it XMM-Newton} EPIC-pn spectrum (observation ID 0840220701), taken simultaneously with the HST COS observations. The model consists of the disc black-body component (\texttt{dbb}, dot-dashed orange curve), Comptonisation (\texttt{comt}, dashed blue line) dominant in the extreme-UV and soft X-rays, power law (\texttt{pow}, dotted green line), cut off on both ends at 13.6 eV and 150 keV, and finally a Gaussian line (\texttt{gaus}, solid purple line) for the 6.4 keV iron feature. The model parameters are given in the text.}
    \label{fig:SED}
\end{figure}
        
To proceed with the modelling of the broad emission lines, we fitted the SED of the incident radiation in SPEX in the following way. At energies covered by the HST observations, corresponding to wavelengths ranging from $1210$ to 8200~\AA, we used several data points extracted from the Galactic extinction-corrected spectra during the emission-line fitting. For the X-ray band, we used the EPIC-pn spectrum from observation ID 0840220701. The global fit was performed using the $\chi^2$ statistic for the UV and optical data points and the $C$-statistic for the X-ray spectrum. The model was constructed as follows.

In the hard X-rays, we fitted the continuum as a power law with a photon index of $\Gamma = 1.66 \pm 0.02$, with an exponential cut-off at 150 keV \citep{Perola2002} and at 13.6 eV at the low-energy end. The iron emission feature with the rest-frame centroid energy of $E_0 = 6.35\pm 0.04 ~\rm keV$ was fitted as a Gaussian line, with a FWHM of $0.3 \pm 0.1~ \rm keV$.

In the soft X-ray band, we used the emission lines and absorption components identified in the RGS spectrum in the model, with parameters fixed to the best-fitting values from the RGS data fitting.
We found the X-ray soft excess was described well with a warm corona Comptonisation model \citep{Done2012, Kubota2018} with a best-fitting optical depth of $\tau = 14 \pm 1$ and an electron temperature of $k_{\mathrm{B}}T_{\mathrm{e}} = 0.28 \pm 0.03~\rm keV$. The seed photon temperature was tied to that of a disc black-body component (\texttt{dbb} in SPEX), which we used for the UV and optical AGN continuum emission. 

Besides the Galactic extinction described in Sect. \ref{ssec:UVoptdata}, an additional extinction correction was necessary to recover a realistic UV continuum profile. However, due to the absence of signal in the extreme--UV band, it was not possible to simultaneously constrain the \texttt{dbb} temperature and the colour excess of the host galaxy extinction. Therefore, we tied the disc black-body temperature to 10 eV, with which the fit yielded an extinction correction described with $E(B-V) = 0.032\pm 0.005$. 

Finally, we added a stellar light component to account for the host galaxy contribution to the spectrum. Assuming the majority of the stellar emission comes from the central region of the galaxy, given the slit (52X0.2) aperture, we adopted the bulge template of \citet{Kinney1996}.

The resulting BLR-ionising SED is presented in Fig. \ref{fig:SED}. The logarithm of its bolometric luminosity in $\mathrm{erg\,s^{-1}}$ is $\log L_{\mathrm{bol}} = 44.90$, and the ionising luminosity, defined as the luminosity between 1 and 1000 Ryd, is $\log L_{\mathrm{ion}} = 44.58$.

\subsection{The LOC model}\label{ssec:locdefinition}

The LOC model assumes that the (broad-)line-emitting gas forms a complex environment that spans over a large range of densities and radial distances to the ionising source. With that, the total line luminosity can be obtained as an integral of density- and distance-dependent contributions $L(n, r)$, that is

\begin{equation}\label{eq:loc}
L_{\rm line} \propto \iint L(n, r) ~\psi(n, r) ~\mathrm{d}n ~\mathrm{d}r.
\end{equation}

\noindent For simplicity, we assume the cloud property distribution function $\psi(n, r)$ is separable, $\psi(n, r) = f(r)g(n)$. The distribution function in $r$ then represents the differential radial cloud covering fraction, for which we assume $f(r)\propto r^{\gamma}$, which was proven to be satisfactory by for example \citet{Baldwin1995} and \citet{Ferguson1997}. Similarly, we put $g(n) \propto n^{\beta}$, setting $\beta = -1$, which has been shown to be applicable to quasars and Seyfert~1 AGNs in the LOC framework \citep[e.g.][]{Baldwin1995, Korista2000, Costantini2007}. The assumption of this power-law density distribution is also supported by results from magnetohydrodynamic simulations presented by \citet{Krause2012}. The density throughout the individual clouds is assumed to be constant.

Furthermore, physically motivated constraints on the upper and lower bounds of the density distribution can be placed as follows. At low densities, below $\sim\! 10^8\rm ~cm^{-3}$, the conditions in the gas would lead to emission through forbidden transitions, which is absent in the data. At the same time, the gas with high density, that is, above $\sim\! 10^{13}\rm ~cm^{-3}$, is unlikely to contribute significantly to the line emission because of line thermalisation, forming a natural upper bound to the model density distribution.

For the radial dimension of the model, we assume the BLR to be confined somewhere between $10^{14.5}\rm~cm$ ---which corresponds to one hundred times the gravitational radius--- and, as proposed by \citet{Netzer1993}, the dust sublimation radius. The latter can be estimated from the AGN bolometric luminosity to $10^{17.6}-10^{18.0}\rm~cm$ in this source, depending on the dust composition \citep[see][for more details]{MorNetzer2012}. We adopt the larger value as an upper limit for the extent of the photoionised gas; we address this choice in Sect. \ref{sec:discussion}.

In addition to the above-mentioned criteria for the gas properties, we exclude gas with ionisation that is too high to efficiently form X-ray lines. In particular, we require the product of the ionisation parameter $U$ \citep{Davidson1977} and the speed of light $c$ in $\rm cm\,s^{-1}$, $\log(Uc) \leq 11.25$ as in \citet{Korista2004}. This condition is equivalent to $ \log \xi \leq 2.34 $\footnote{$U$ is defined as the dimensionless ratio of the flux of hydrogen-ionising photons to the product of hydrogen density and the speed of light. As such, the relation between $\log \xi$ and $\log U$ is dependent on the SED of the ionising radiation. In this case, $\log U = \log \xi - 1.57$.}. At the other end of the model grid, at high gas densities and low ionising-photon fluxes, we impose a lower limit of $\log U > -5$.

To constrain the radial distribution of the gas using the LOC approach, we sampled the density--radius parameter space with a step of 0.125 dex, creating a grid of photoionisation models with \textsc{Cloudy} for a gas of given properties and subjected to the ionising radiation described in Sect. \ref{sec:SED}. We assumed a total hydrogen column density of $10^{23}~\rm cm^{-2}$ and the \textsc{Cloudy} default solar elemental abundances. An example of the predicted equivalent widths as a function of hydrogen particle density and ionising flux is illustrated in \ref{fig:W-phi-n}. To finalise the model preparation, we extracted the predicted line luminosities for each point of the density--radius grid, including the weaker lines that may blend with the observed line profile.

\subsection{Broad-band LOC fitting}\label{sec:fitting}
\subsubsection{The baseline model: A single LOC component}\label{ssec:1loc}

\begin{figure*}[ht]
    \centering
    \resizebox{0.85\hsize}{!}{\includegraphics{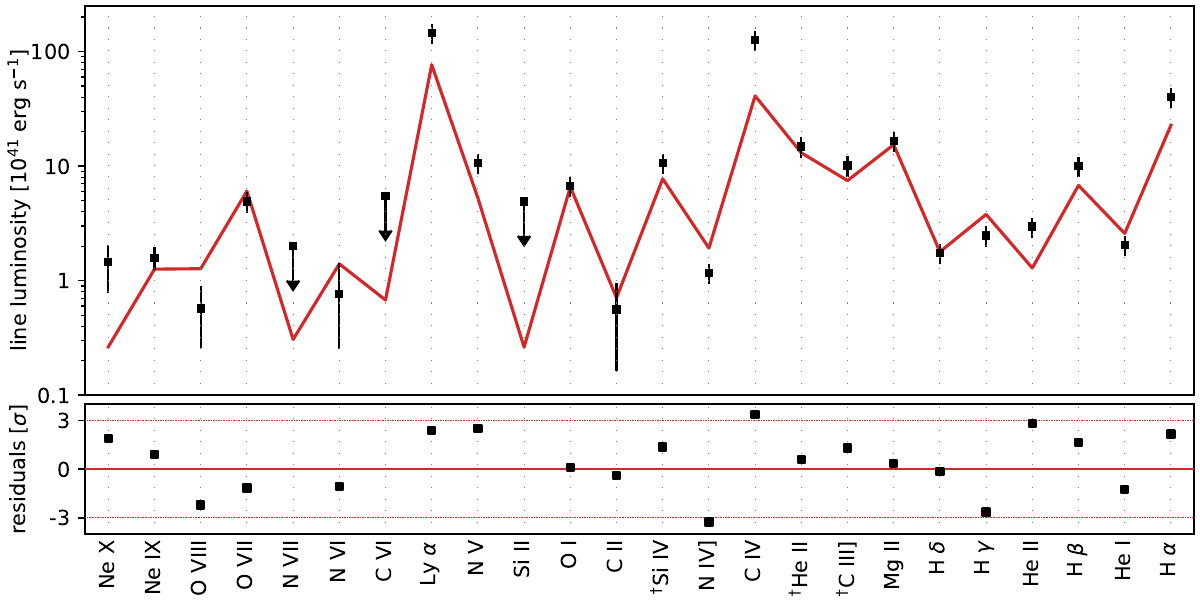}}
    \caption{Luminosities of the X-ray, UV, and optical lines and the best-fitting baseline LOC model. The data points displayed in the upper panel represent the intermediate+broad components of the observed line luminosities corrected for Galactic extinction and ordered by wavelength (increasing to the right). The spectral lines labelled with a dagger, namely \ion{Si}{iv}, \ion{He}{ii}, and \ion{C}{iii],} represent summed luminosities from blended nearby lines, as explained in Table \ref{tab:lines-UVoptical}. The upper limits were not taken into consideration during fitting and are only plotted for reference. In the bottom panel, the fit residuals for both models are plotted for comparison.}
    \label{fig:1loc_g1}
\end{figure*}

To test whether the LOC model can sufficiently describe the broad line emission, we started with a simple scenario, designed as follows. We decided to limit our fitting to the components classified as intermediate and broad in Table \ref{tab:lines-UVoptical}. The reason for this step is that the detection of the very broad lines is limited to the strongest ones in our sample and that these components are highly redshifted, which makes their inclusion in the fitting problematic. As the large width is accompanied by rather small amplitude relative to other components, it is possible that, if present also in weaker lines, this shallow part becomes undetectable, blending into the continuum. Adding the very broad component in any case led to a considerably poorer fit statistic. We note that this filtering did not concern the X-ray lines, where only one Gaussian component was sufficient to describe the observed broad lines.

To probe the distance and radial extent of the line-emitting region, we fixed the radial distribution power-law slope to $-1$, with which the cloud properties are the most broadly distributed over the full radial range, as favoured by the LOC basic assumptions. The integrated global covering fraction $C_{\mathrm{f}}$ was allowed to vary in the range $0.05-0.6$. This range encompasses previous estimates of the BLR covering factor \citep[see][]{Costantini2016}. For clarity, we present all parameters related to the baseline model in Table \ref{tab:LOCpars}.

\begin{table}[ht]
\caption{Baseline LOC model-related parameters.}
\centering\label{tab:LOCpars}
\begin{tabular}{l|c|c}
\hline\hline
Parameter & Value/allowed range & Status \\
\hline
$\log (n/\mathrm{cm^{-3}})$ & $\left[8.0; 13.0\right]$ & fixed \\
$\log (r_{\mathrm{min}}/\mathrm{cm})$ & $\left[14.5; R_{\mathrm{dust}}-1\right]$ & free \\
$\log (r_{\mathrm{max}}/\mathrm{cm})$ & $\left[14.5+1; R_{\mathrm{dust}}\right]$ & free \\
$\beta$ & $-1$ & fixed \\
$\gamma$ & $-1$ & fixed \\
$C_{\mathrm{f}}$ & $\left[0.05; 0.6\right]$ & free \\
$\log (N/\mathrm{cm^{-2}})$ & 23.0 & fixed \\
abundances & solar & fixed \\
$\log U_{\mathrm{min}}$ & $-5$ & fixed \\
$\log U_{\mathrm{max}}$ & $0.77$ & fixed \\
$\log (L_{\mathrm{ion}}/\mathrm{erg\,s^{-1}})$ & $44.58$ & fixed \\
\hline
\end{tabular}
\end{table}

The model that yielded the best-fit statistic amongst all possible combinations of inner and outer radius has the following best-fitting parameters. The inner radius is at $\sim10^{15.3}~\rm cm$, which is  larger than the innermost point of the grid by almost an order of
magnitude, and the outer edge is at $\sim 10^{17.6}~\rm cm$, close to the dust-sublimation radius. The global covering fraction is $C_{\rm f} = 0.52\pm 0.05$. 

In the top panel of Fig. \ref{fig:1loc_g1}, the model is plotted along with the data points. It can be seen that, while a statistically good fit cannot be obtained, the predicted UV and optical line luminosities generally agree with the observed ones within 3.5$\sigma$ uncertainty. However, the brightest hydrogen lines, \ion{C}{IV} and \ion{N}{V} are systematically underpredicted by the model. 

In the X-ray band, the LOC model predicts line production in relatively more weakly ionised gas, namely \ion{Ne}{IX}, \ion{O}{VII,} and \ion{N}{VI}, and can fully account for this observed X-ray emission. The only well-constrained line representing more highly ionised gas, \ion{Ne}{X}, as well as the upper limits of \ion{C}{IV} and \ion{N}{VII,} are underpredicted by nearly an order of magnitude. This suggests that an emitter in a higher ionisation state is responsible for these lines, and is likely present closer to the source of the ionising radiation. However, the models reaching down to a smaller inner radius of the BLR clouds provide a less accurate description of the data overall. Assuming the X-ray emission could indeed originate from the BLR gas, this outcome may indicate that the broad-line-contributing clouds have a more complex distribution. 

It is important to note that the exclusion of the X-ray lines does not significantly alter the model parameters. The best-fitting model values of the case where only the UV and optical lines are used in the fitting are therefore nearly identical to those in Fig. \ref{fig:1loc_g1}.

In summary, while the best-fitting model captures the general properties of the data, some important aspects need to be investigated further. In the following, we explore effects that may influence the BLR emission, including the consequences of the assumptions associated with the baseline model construction.

\begin{table*}[ht]
\caption{Best-fitting model properties.}
\centering\label{tab:LOCfit}
\begin{tabular}{l|cccc|cccc|c}
\hline\hline
\multirow{2}{*}{Model} & \multicolumn{4}{c|}{Component 1 ($\log U < 0.77$)} & \multicolumn{4}{c|}{Component 2 ($\log U < 2.0$)} & \multirow{2}{*}{$\chi^2/\mathrm{d.o.f.}$}\\
 & $\log r_{\mathrm{in}}$ & $\log r_{\mathrm{out}}$ & $\gamma$ & $C_V$ & $\log r_{\mathrm{in}}$ & $\log r_{\mathrm{out}}$ & $\gamma$ & $C_V$ &  \\
 
 \hline
baseline      & 15.250 & 17.625 & $-1$            & $0.52\pm 0.05$ & \multicolumn{4}{c|}{$\cdots$} & $71.6/18$ \\
$\gamma$ free & 16.000 & 18.000 & $-1.6\pm 0.1$  & $0.56\pm 0.08$ & \multicolumn{4}{c|}{$\cdots$} & $59.7/17$ \\
2 comp.       & 16.000 & 18.000 & $-1$            & $0.17\pm 0.06$ & 16.000 & 17.000 & $-1\pm 1$ & $0.3\pm 0.1 $ & $61.1/14$ \\
\hline
\end{tabular}
\tablefoot{The baseline model from Sect. \ref{ssec:1loc} is presented in the first line, followed by a model with an additional free parameter ($\gamma$, see Sect. \ref{ssec:gamma}) and a two-component model from Sect. \ref{ssec:2loc}.}
\end{table*}

\subsubsection{Intrinsic dust extinction}\label{ssec:extinction}

\begin{figure}
    \centering
    \resizebox{\hsize}{!}{\includegraphics{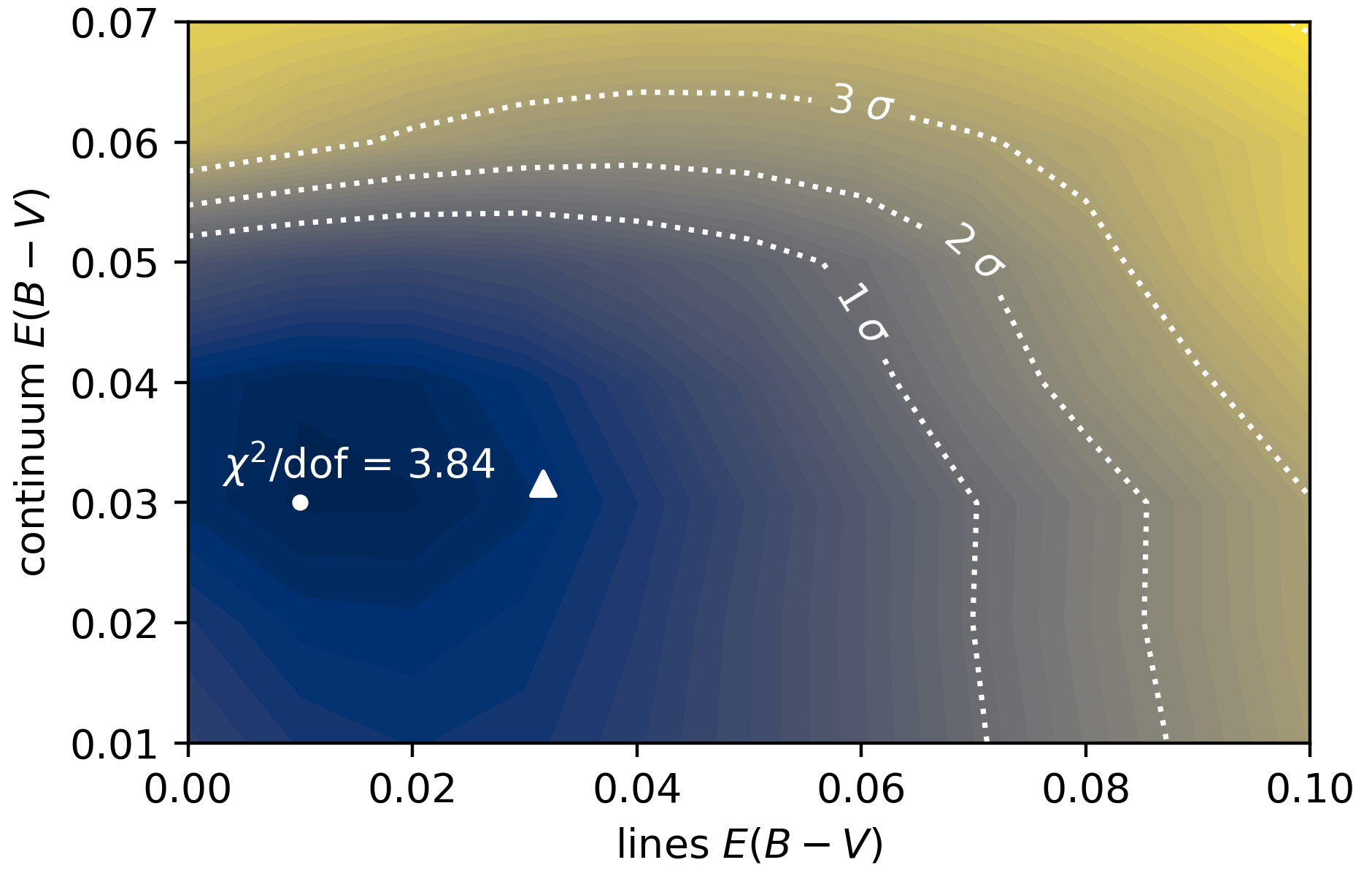}}
    \caption{Best-fitting reduced $\chi^2$ (for 18 degrees of freedom) of the single-component LOC model fits as a function of SED- and line-affecting extinction intrinsic to Mrk\,110. The contours signify the 1, 2, and 3 $\sigma$ confidence regions with respect to the best-fit statistic (marked with a dot). The triangle represents the intrinsic extinction considered in the baseline model ($E(B-V) = 0.032$, for both the lines and the ionising SED).}
    \label{fig:ext-maps}
\end{figure}

The global fitting of the SED used in the photoionisation modelling yielded additional dust extinction associated with
the host, with $E(B-V) = 0.032\pm 0.005$. However, this result is dependent on the assumptions made about the UV--optical continuum properties ---however reasonable--- because of the intrinsically limited information from the extreme-UV band. Therefore, we examine whether a significantly different amount of extinction could have affected the observations.

Substantial internal reddening associated with the Mrk\,110 AGN has been reported in the literature, with $E(B-V)$ in the range of 0.23-0.63 \citep{Grupe2010, Winter2010, Jaffarian2020}. These values were derived from the Balmer ratio, which is typically used as an extinction indicator. Indeed, its value derived from our data is also rather large \ion{H}{$\alpha$}/\ion{H}{$\beta$} = 4.8, suggesting the emission is affected by intrinsic reddening of $E(B-V) \approx 0.4$. This estimate is made using the \citet{Gordon2003} SMC Bar Average extinction curve with $R_V = 2.74$ and the intrinsic line ratio of 3.06 proposed for BLR gas in a sample study of Seyfert 1 and quasar spectra by \citet{Dong2008}.

However, the BLR gas spans a large range of physical parameters, including those in which the line ratio is far from the standard Balmer decrement. As a consequence, the observed \ion{H}{$\alpha$}/\ion{H}{$\beta$} is
also dependent on the gas distribution in density and ionisation state, making it an unreliable extinction indicator, as was found to be the case for Mrk\,110 by \citet{Bischoff1999} and, more recently, \citet{Homan2022}. 

A better, yet still not  optimal, indicator is the ratio of \ion{He}{ii} 1640 and \ion{He}{ii} 4686 luminosities. Taking into account the uncertainties involving blending with neighbouring lines and the intrinsic value of the ratio \citep[between 7 and 9;][]{Bottorff2002}, the expected reddening corresponds to $E(B-V) \approx 0.03$. While this value is consistent with the results from the SED used so far, we nevertheless examine a range of acceptable values of the colour excess in our modelling.

Specifically, we reconstructed the incident ionising SED by fitting the HST continuum data points with the same model as described in Sect. \ref{sec:SED}, but keeping the colour excess of the extinction model at a fixed value. With the degeneracy between the parameters, having the disc black body temperature and normalisation free allowed us to obtain a satisfactory fit for a range of extinctions, namely $0.01 \leq E(B-V)_{\rm SED} \leq 0.07$ (with a step of 0.01), using the \citet{Gordon2003} SMC Bar Average extinction law with $R_V = 2.74$. As a consequence, the resulting SEDs (see Figure \ref{fig:extSED}) span nearly an order of magnitude in the ionising luminosity, specifically $10^{44.5}-10^{45.3}~\mathrm{erg\,s^{-1}} $.

With the photoinisation models derived for these SEDs, we tested the impact of the different recovered SED shapes on the line luminosity fitting. As it is physically possible for the continuum illuminating the BLR and the BLR emission itself to be affected by a different amount of reddening in our line of sight, we examined a wider range of extinction for the line corrections, $0.0 \leq E(B-V)_{\rm lines} \leq 0.1$, and allowed for all possible combinations of extinction corrections applied to the data and the ionising continuum. We note that the upper limit imposed on the outer radius in these fits changes according to the dust-sublimation radius, which, in turn, is  dependent on the bolometric luminosity of the SED used for the photoionisation modelling.

The best-fitting reduced $\chi^2$ values and contours signifying commonly used confidence intervals are plotted for the examined grid of extinction parameters in Fig. \ref{fig:ext-maps}. It is immediately noticeable that while the model is somewhat less sensitive to line-affecting extinction corrections (suggesting $E(B-V)_{\rm lines} \lesssim 0.1$), models with SEDs derived for $E(B-V)_{\rm SED} \lesssim 0.06$ are preferred at $3\sigma$ significance. The minimal fit statistic is achieved for conditions close to those considered so far ($E(B-V) = 0.032$), with $E(B-V)_{\rm lines} = 0.01$ and $E(B-V)_{\rm SED} = 0.03$, but the fit quality is not significantly better. With this result of the extinction parameter space exploration, we continue our analysis with the originally presented UV--optical SED properties for the rest of the paper.

\subsubsection{SED variability}\label{ssec:varSED}

As a NLS1, Mrk\,110 is a variable source \citep[e.g.][]{Peterson1998, Bischoff1999, Homan2022, Vincentelli2021}. The continuum level variations in this source on timescales of days to months can reach a factor of a few in the X-ray, UV, and optical bands. The variability on these timescales is relevant for the spectral signatures of the BLR due to its radial extent, as the gas at each distance then effectively `sees' a different ionising radiation at any given time. It is therefore important to assess whether these effects can play an important role in our analysis.

Multi-band \textit{Swift} \citep{Gehrels2004} observations taken prior to our HST and {\it XMM-Newton} observations allowed us to constrain the X-ray, UV, and optical variability during the preceding 24 days. Specifically, we extracted light curves of the source in several energy bands using the X-ray Telescope \citep[XRT,][]{Burrows2005} and UV-Optical Telescope \citep[UVOT,][]{Roming2005} on board. The light curves were extracted from UVW1, UVW2, UVM2, U, B, and V filters, and $0.3-1.5$ keV and $1.5-10$ keV bands. Based on these data, the optical and UV flux did not change by more than 10\%. In the X-ray bands, the flux did not change by more than 25\%\ and overall, the spectral shape remained unchanged.

Assuming a maximum change of 25\%\ of the X-ray flux, we tested how the predictions for the BLR properties would be altered under the different SED. Namely, we changed the normalisation of the X-ray-affecting continuum components (see Sect. \ref{sec:SED}) by a factor of 0.75 and 1.25, respectively, simply parametrising two scenarios: a uniform decrease and increase in the X-ray flux.

As expected, the model grid predictions for the X-ray band are the ones most affected by variability. However, with small changes to the model parameters, a fit of equivalent statistical properties can be achieved with either of the scenarios. With a maximum variation of 25\%, the best-fitting models do not change within the errors of the brightest X-ray lines (\ion{O}{vii} and \ion{Ne}{IX}). We note that the \ion{Ne}{x} line, which is possibly more sensitive to a higher X-ray flux, is not better fitted if a different SED is adopted. However, we cannot completely rule out that a sudden flare, just before our observation, could have produced more highly ionised X-ray lines, including \ion{Ne}{x}. 

\subsubsection{Slope of the radial distribution}\label{ssec:gamma}

\begin{figure}[ht]
    \centering
    \resizebox{\hsize}{!}{\includegraphics{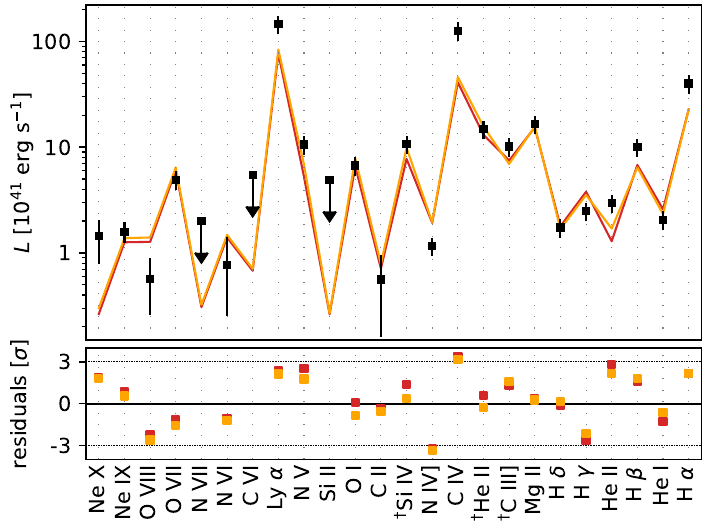}}
    \caption{Luminosities of the X-ray, UV, and optical lines fitted with LOC models. \textit{Top:} Best-fitting single-component LOC model with the radial distribution power-law index $\gamma$ considered as a free parameter in the fit (orange; see Sect. \ref{ssec:gamma} for details). \textit{Bottom:} Corresponding fit residuals in units of the uncertainty $\sigma$ of the line luminosity measurements. The baseline model (with the radial distribution power-law index fixed to $\gamma = -1$; see Sect. \ref{ssec:1loc} and Fig. \ref{fig:1loc_g1}) and the corresponding residuals (red) are displayed for reference. The data are displayed as in Fig. \ref{fig:1loc_g1}.}
    \label{fig:1loc_gfree}

    \resizebox{\hsize}{!}{\includegraphics{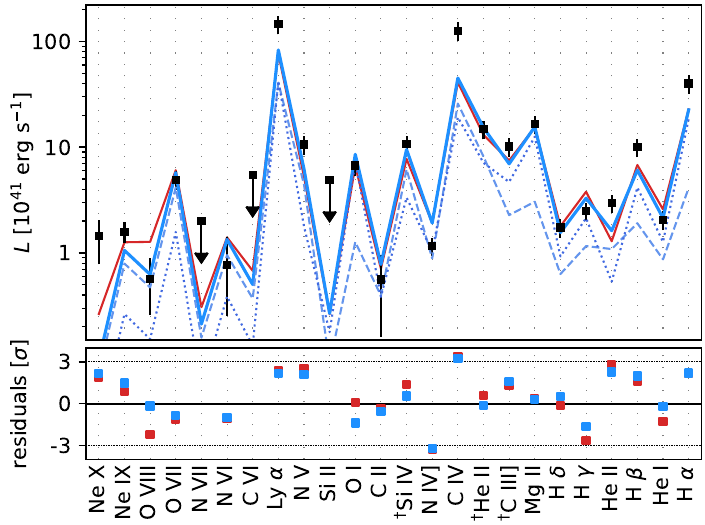}}
    \caption{Luminosities of the X-ray, UV, and optical lines fitted with the LOC models. \textit{Top:} Best-fitting two-component LOC model (solid blue line; see Sect. \ref{ssec:2loc} for details). The individual model components are also given: with a dotted blue line there is the component with $\gamma$ fixed to $-1$ and the dashed blue line represents the component with $\gamma$ considered as a free parameter. \textit{Bottom:} Corresponding fit residuals in units of the uncertainty $\sigma$ of the line-luminosity measurements. The baseline model (with the radial distribution power-law index fixed to $\gamma = -1$; see Sect. \ref{ssec:1loc} and Fig. \ref{fig:1loc_g1}) and the corresponding residuals (red) are displayed for reference. The data are displayed as in Fig. \ref{fig:1loc_g1}.}
    \label{fig:2loc}
\end{figure}

As the best-fitting models presented so far have failed to cover the entire range of observed line luminosities, the properties of the cloud distribution should be addressed in more detail. The first parameter to examine is the power-law index of the radial distribution differential covering fraction. In the above, the value adopted, $\gamma = -1$, was kept fixed. However, it might be sensible to leave this value to be determined in the fitting procedure, provided the resulting value remains reasonably close to $-1$. This condition originates in the general requirement of the BLR model, that is, a broad coverage of the density--radius parameter space.

With this additional parameter free, a noticeably better fit is obtained (see the top panel of Fig. \ref{fig:1loc_gfree}, and Table \ref{tab:LOCfit} for an overview of the parameters). To assess the best-fitting model performance with respect to the case with $\gamma = -1$, we employed Akaike's information criterion \citep[AIC;][]{Akaike1974}, which allows the comparison of non-nested models. For a model with $k$ free parameters and the maximum likelihood value $\mathcal{L}_{\mathrm{max}}$, the AIC is defined as

\begin{equation}
  \mathrm{AIC} = 2k - 2 \ln(\mathcal{L}_{\mathrm{max}}).  
\end{equation}

\noindent The difference in the AIC values between the model with the lowest AIC value, $\rm AIC_{min}$, and a competing model $i$, $\mathrm{AIC_{min}} < \mathrm{AIC}_{i} $, then determines whether it is plausible that the model $i$ is the best, given the data. Interested only in $\Delta \rm AIC$, we can write directly $\chi^2 = -2\ln \mathcal{L}$.

Considering this criterion alone, the best-fitting model with $\gamma$ left free provides a significantly better description of the data. With $\Delta \rm AIC = 10.0$, there is little empirical evidence supporting the scenario with $\gamma $ fixed to $-1$. However, the $\chi^2$ statistic remains relatively high, as the previously identified strong outliers are still present. Furthermore, the slope of the radial distribution, $\gamma = -1.6\pm 0.1$, is rather steep. As a result, the bulk of the modelled clouds facing the ionising radiation is located close to the inner radius. Moreover, the line-emitting region necessary to recover the observed line luminosities extends out to the upper limit of the permitted radial extent ---the dust sublimation radius, beyond which little photoionisation can occur. As the steep distribution suggests a departure from the premise of the LOC model, that is, broadly distributed cloud properties, other scenarios that could explain the BLR emission ought to be examined.

\subsubsection{Two LOC components}\label{ssec:2loc}

In order to examine whether a more complex model could provide a better fit to the data, we added another LOC component to the fitting procedure. As for the previous model, we allowed the slope of the new radial distribution to vary, along with the global covering fractions of both components. To facilitate a better match to the observed \ion{Ne}{x}, we reintroduced a part of our model grid with more highly ionised gas for the new component. Namely, we included predictions for gas with ionisation $\log U \leq 2 $ (or $\log \xi = 3.57$). The effect of this change on the model grid can be seen in Fig. \ref{fig:W-phi-n}, where the equivalent width of the X-ray lines well detected in Mrk\,110 are plotted as a function of the hydrogen-ionising photon flux and the gas density. To restrict the fitting to physically meaningful configurations, we limit the inner radius of this additional LOC component to below or at the inner radius of the less ionised one.

The fitted composite model and its components are plotted in Fig. \ref{fig:2loc}. While the model as a whole provides a slightly better match to the data, in particular the less luminous lines, the overall quality of the fit is not improved with respect to the single-component scenario, given that four more free parameters were added. Also, as can be seen in Table \ref{tab:LOCfit}, the power-law index of the second distribution is also not well constrained. Nevertheless, it can be seen that this model also prefers a configuration that enhances the contribution from the inner part of the radial extent of the modelled clouds. Again, only a minimal improvement is seen in the case of the systematically underpredicted \ion{C}{iv} and bright hydrogen lines.

\section{Discussion}\label{sec:discussion}
\subsection{The multi-band BLR of a NLS1}

The X-ray-to-optical broad lines of Mrk\,110, spanning over two orders of magnitude in luminosity, can be reasonably well described with a single-component LOC model. Given the simplicity of the basic assumptions of the method, valuable information can be obtained from the results, despite the presence of several outliers standing out in the analysis, which are addresses in the following subsections.

The best-fitting standard LOC model predicts that the gas will lie at distances of $ 15.3 \lesssim \log r \mathrm{[cm]} \lesssim 17.6$. This radial extent is consistent with the reverberation mapping analysis of \citet{Kollatschny2001}, which focuses on the optical band lines, where the measured lags correspond to the range of $10^{15.9}-10^{17.0}~\rm cm$. The outer edge is consistent with the expected radius at which the dust grains can survive \citep{MorNetzer2012}, assuming the medium is exposed to the same incident continuum as at the inner edge of the BLR. For the upper and lower limits on radius, a similar result is obtained if the distribution of the modelled gas properties is allowed to depart from the standard, most simple solution. This was achieved by freeing one of the parameters  in
the fitting procedure, namely the radial distribution power-law index, and in a second instance by the addition of another LOC component. While these more complex scenarios give an overall better match to the observed line luminosities, it needs to be stressed that caution is necessary when assessing the achieved improvement.

Most importantly, the predictions for the brightest lines ---which are systematically underpredicted by the model--- remain unchanged. As a result, the improvement of the fit statistic is minimal given the number of added free parameters of the two-component model. For the models to reach a significantly more accurate description of the data, a larger outer radius of the model grid would be necessary. However, in our analysis,  the outer radius of the model grid is limited by the dust sublimation radius of silicate grains, at approximately $10^{18}~\rm cm$. Even there, as mentioned above, dust can, in principle, be present, as the dust sublimation of purely graphite dust grains occurs at a higher temperature (corresponding to 0.4~dex smaller radius).

Generally, the inclusion of gas exposed to higher ionising fluxes (at radii below $ 10^{15.3}~\mathrm{cm}$) in the model is detrimental to the overall fit quality. Considering the best-fitting model results, which favour gas produced between $ 10^{16}$ and $10^{18}~\mathrm{cm}$, the BLR is about an order of magnitude more compact in comparison with other objects for which the same modelling was used. This agrees with the relatively small width of the emission line profiles, assuming the broadening results mostly from Keplerian motion around the central supermassive black hole. For reference, the FWHM of the broad component of \ion{H}{$\beta$}, of namely $4400~\rm km\,s^{-1}$, corresponds to a distance of $10^{16.1}~\rm cm$ if Keplerian motion around a supermassive black hole with a mass of $2\times 10^7 \, M_{\odot}$ is assumed.

The slope of the density distribution, $\beta = -1$, adopted here proves to be a suitable choice, which is also the case of the broad-line Seyfert~1 AGNs mentioned above. We tested different values for this slope with all considered model scenarios, ranging from $-0.7$ to $-1.2$, with a step of 0.1, and found no significant improvement of the fit statistic.

The relatively high integrated global covering fraction $C_{\mathrm{f}}$ is also found in other objects; for example \object{NGC~5548} \citep{Korista2000} or Mrk~509 \citep{Costantini2016}. Together with the absence of broad absorption line features, this high fraction suggests a BLR geometry where the line-emitting gas is located away from the line of sight, leaving at least an opening angle consistent with the AGN inclination  empty, assuming axial symmetry.

Compared to normal Seyfert~1 AGN, the power-law index of the radial distribution of the model is considerably steeper, with $\gamma \approx -1.6 $ for the  presented best-fitting model, as opposed to $ -1.2$ in NGC~5548 \citep{Korista2000}, $-1.0$ in \object{Mrk\,279} \citep{Costantini2007}, or \object{Mrk\,509} with $\gamma \approx -1.1$ \citep{Costantini2016}. In contrast, the steep slope is consistent with the findings of \citet{Nagao2006} for composite quasar spectra (sampled from objects at $2.0 \leq z \leq 4.5$ and with the $B$-band absolute magnitude in the range $-24.5 \ge M_B \ge -29.5$), finding that $-2.0 < \gamma < -1.5$. The steep radial distribution enhances the importance of more highly ionised regions, located closer in. Nevertheless, with the density distribution taken into account, most of the emission in the strongest lines still originates from low-density clouds at large distances. Finally, while the power-law index of $-1.6$ yields a significantly better fit, the fact that the value departs quite far from $-1$ alludes to the presence of additional processes acting beyond the basic premise of the LOC approach.

\subsubsection{Complexity of the BLR}
Despite the relatively good match to the data, given the large range in ionisation and luminosity, the LOC model fails to capture the full extent of the detected line luminosities, underpredicting mainly the brightest hydrogen lines (\ion{Ly}{$\alpha$} and \ion{H}{$\alpha$}) and the \ion{C}{IV} doublet (Fig. \ref{fig:1loc_g1}). Because adjustments to the line-emitting region parameters within the model framework do not lead to a sufficient improvement of the fit, more complex scenarios need to be considered. A possible explanation might lie in the intrinsic simplicity of the model, which does not cover effects such as gas clumping \citep{Waters2021} or self-shielding by the line-emitting clouds \citep{Matthews2020}, although with the latter, it should be kept in mind that the model is not very sensitive to differences between the real and assumed SED `seen' by the clouds \citep[e.g.][or see Sect. \ref{ssec:varSED}]{Baldwin1997}. Moreover, the constant global covering factor or the power-law distributions assumed for the density and distance distributions might be an oversimplification. 

In addition to the aspects related to geometrical structure, elemental abundances of the gas may play an important role as well. However, it is difficult to obtain reliable constraints for the chemical composition from the observed data without introducing, in turn, strong assumptions regarding the gas properties. A detailed analysis of the chemical composition of the BLR gas in an attempt to overcome such issues is beyond the scope of this study. 

Despite several studies \citep{Jaffarian2020, Grupe2010, Winter2010} suggesting there is substantial internal reddening (with a colour excess of as large as $E(B-V) \approx 0.4$), which is manifested through a large (broad-line) Balmer ratio, the LOC model disfavours such an explanation. Specifically, the physical conditions in the BLR gas do not necessarily lead to the traditionally adopted value of this indicator assumed for gas unaffected by extinction. Furthermore, we showed that significantly poorer fits are achieved when extinction exceeding $E(B-V)\gtrsim 0.06$ is considered. This result suggests that indeed a smaller amount of dust is present in this AGN and that this may also be case in other systems where the Balmer decrement appears unusually large.

Importantly, the gas producing UV lines also comprises a significant source of the detected X-ray emission, in particular the lower-ionisation lines (\ion{N}{vi}, \ion{O}{vii}, \ion{Ne}{ix}) present in this source. Recently, the \ion{O}{VII} flux was found to be strongly linearly correlated with the monochromatic continuum flux at 0.5 keV \citep{Reeves2021}, suggesting that the source of the X-ray line is in the accretion disc. However, taking into account the neutral and ionised absorption affecting the flux of the \ion{O}{vii} line, we find that the \ion{O}{vii} flux is consistent with being produced within the BLR.

The relatively large amount of detected \ion{Ne}{X} emission, in contrast with the \ion{O}{VIII} line, is indicative of its origin in a different, even more highly ionised region present in the AGN. A simple explanation would be its origin in the accretion disc, even closer to the central engine, but it could also be produced in a highly ionised layer of the torus \citep{Bianchi2005, Costantini2010, Ponti2013}.

\subsubsection{Possible additional wind component}\label{sec:wind}

While the LOC model is capable of explaining the general similarity between the BLR spectra seen in many objects, including Mrk\,110, the details of the kinematic and geometrical structure remain unconstrained with this approach. Furthermore, the systematic redshift of the centroids of the broadest line components in Mrk\,110, which are prominent in lines of a wide range of ionisation, from the bright X-ray lines to the \ion{C}{iv} doublet and also \ion{H}{$\alpha$}, deserve more attention.

Recently, \citet{Matthews2020} presented a model of the BLR in the form of a clumpy biconical disc wind, aiming to recover the observable spectral properties through radiative transfer simulations in the wind-like environment, and also presented an alternative to the static picture of the LOC-type models. As the model proved successful in reproducing the basic properties of the BLR spectral behaviour, the predictions are interesting for the BLR in Mrk\,110 as well. More importantly, other studies of the broad-line emission from this source consider its origin in a disc wind a viable explanation to the observed line properties \citep[e.g.][]{Kollatschny2001, Pancoast2012, Homan2022}.

However, the disc-wind scenario was presented for a quasar-like source, preventing the models of \citeauthor{Matthews2020} from being directly transferable to our data. In particular, their simulations were carried out assuming an SED with a larger relative contribution from the X-ray band, and therefore the resulting line ratios might not accurately represent the situation in Mrk\,110. It is also unsurprising that the modelled line profiles are generally much wider, enhancing the effect of line blending.

With the caveats mentioned above, we illustrate that the inclusion of models with physically motivated wind-like structure in the general LOC scenario can offer valuable information that can be used to improve the description of the BLR. For this exercise, we worked with the synthetic spectra provided by \citeauthor{Matthews2020} as follows. We extracted the predicted line fluxes for all available combinations of the wind launch radii, minimum and maximum opening angles, and volume filling factors. We note that \citeauthor{Matthews2020} define the wind `clumpiness' as the inverse of the gas volume filling factor.

As the predicted spectra depend on the AGN inclination, we fixed the viewing angle to $35^{\circ}$, in agreement with the constraint for Mrk\,110 by \citet{Wu2001}. We note that the fluxes of only the brightest lines could be extracted from the synthetic spectra, namely the blend of \ion{Ly}{$\alpha$} and \ion{N}{v}, \ion{Si}{iv}, \ion{C}{iv}, \ion{C}{iii]}, \ion{Mg}{ii}, the blend of \ion{H}{$\beta$}, and \ion{H}{$\alpha$}.

\begin{figure}[ht]
    \centering
    \resizebox{\hsize}{!}{\includegraphics{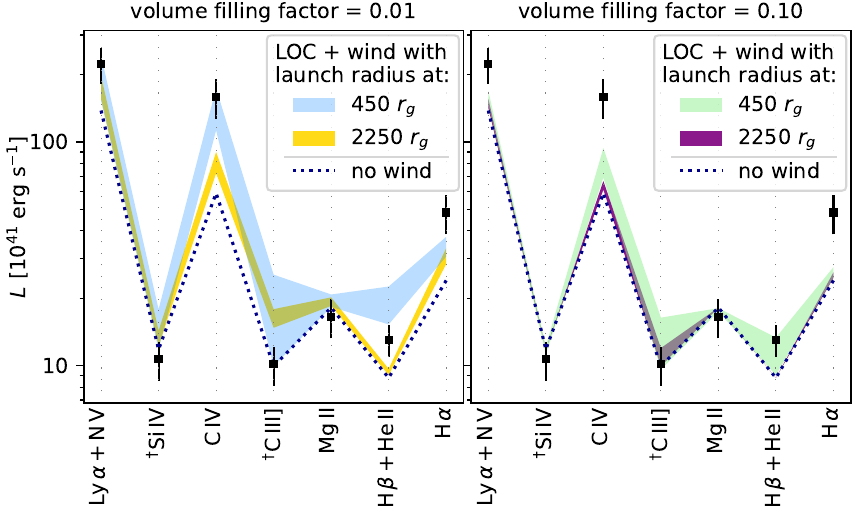}}
    \caption{Sum of the best-fitting LOC model (dotted lines) and the predictions from wind-like BLR models of \citet{Matthews2020} (shaded areas). The wind volume-filling factor of the displayed models is 0.01 (left panel) and 0.1 (right panel). The width of the composite model prediction for each line represents the range of luminosity expected by the wind model assuming different minimal and maximal opening angles of the wind, ranging from nearly axial to nearly equatorial. Following the wind model predictions, we merged the measured luminosities of \ion{Ly}{$\alpha$} with those of \ion{N}{v,} and the
measured luminosities of \ion{H}{$\beta$} with those of \ion{He}{ii}, in addition to other blends (same as in Fig. (\ref{fig:1loc_g1}), marked with a dagger).}
    \label{fig:wind}
\end{figure}

In Fig. \ref{fig:wind}, the wind components are added to the best-fitting single-LOC model described in Sect. \ref{ssec:1loc}. For consistency between the Mrk\,110 data and the wind models, we used here the entire observed broad-line luminosity (by including the broadest emission category of our line classification; Table \ref{tab:lines-UVoptical}). The composite (LOC+wind) models for each launching radius and volume-filling factor are plotted as shaded areas, where the line luminosity range for each line encapsulates different scenarios for opening angles of the wind.
The two panels show the model composites with wind-like components differing by the gas volume filling factor. We note that in the simulations, a higher volume filling factor also means that the wind consists of clumps of lower density and vice versa, affecting the gas properties as a consequence. Each of the panels shows models for winds launched from two distances from the central black hole, 450 and 2250 gravitational radii, respectively. For reference, we note that 450 gravitational radii correspond to $\log r[\mathrm{cm}] = 15.25$ in Mrk\,110, which is a value that is within the same order of magnitude as the inner radius of the best-fitting LOC model. The normalisation of the wind component did not require any adjustments. We note that this is not a formal fit.

While the implications from this exercise are limited by numerous caveats, two elements seem to be significant. Firstly, a generally better representation of the data can be achieved with a smaller launching radius (450 gravitational radii) and a smaller volume-filling factor. Secondly, the most distinct aspect of the models is the behaviour of the blend of \ion{H}{$\beta$} and \ion{He}{II}. This feature also seems to be more sensitive to the minimal and maximal opening angles of the winds. We note that the best match to the data can be achieved with a nearly equatorial wind, with minimal and maximal opening angles of $70\degr$ and  $85\degr$, respectively, measured from the disc normal. This applies for both volume-filling factors shown, namely 0.1 and 0.01. This model geometry, relative to the other
wind configurations, results in little \ion{C}{iii]} and \ion{Mg}{ii} emission, in contrast with the higher ionisation lines. While it is not possible to draw any conclusions or confidently exclude other geometries based on the exercise presented here, the nearly disc-like structure, assuming it is representative of the global BLR geometry in Mrk\,110, is consistent with conclusions from the line-profile-sensitive reverberation mapping study of \citet{Kollatschny2001}.

Generally, while the predicted line ratios for the wind models cannot describe the BLR emission as a whole, their addition is beneficial, contributing as much as 50\,\% of the total \ion{Ly}{$\alpha$} flux. The combination with the likely isotropic LOC-like structure extending to larger radii can then provide a good match to the line luminosities. Furthermore, the presence of an additional wind-like structure located at the inner edge of the BLR can offer an explanation for the systematic redshift of the broadest line components (Sect. \ref{sec:analysis}), suggesting a geometrical factor could be at play. 

\subsection{The X-ray--UV absorbed spectrum}

As mentioned in Sect. \ref{sec:analysis}, two ionised absorption components are detected in the RGS spectrum. The model for the $\log\xi=0.6$ component predicts significant absorption in the \ion{Ly}{$\alpha$} region, specifically in the COS band. Considering the outflow velocity ($2600\pm200~\mathrm{km\,s^{-1}}$), \ion{Ly}{$\alpha$} absorption should be at $1247.7\pm0.8$~\AA\ in the observer's frame of reference. We indeed find three candidates, at 1245.2, 1245.4, and 1246.7~\AA, broadly consistent with the X-ray prediction.

However, a considerable discrepancy is in the equivalent width of the UV absorption line. The X-ray model predicts a \ion{H}{i} column density of $(1.0\pm0.5)\times 10^{15}\,\mathrm{cm^{-2}}$, leading to an expected equivalent width of approximately 2~\AA, while the sum of the observed equivalent widths of all three absorption features is only 0.22~\AA, suggesting the column density is more than an order of magnitude smaller. Nonetheless, there are no other absorption features at the expected wavelengths, meaning that these lines are the only candidates for the UV counterpart of the X-ray warm absorber. Additionally, no other UV absorption lines are observed at this blueshift, as expected given the ionisation of the gas.

A possible explanation for the mismatch between the expected and observed line properties ---which is also noticeable in other objects--- was proposed by \citet{Arav2002}. Among others, the extrapolation of the absorption properties from the X-ray to the UV part of the spectrum relies on the assumption that the outflow covers both the X-ray and UV emission uniformly. Supported by observational evidence, \citeauthor{Arav2002} demonstrated that this assumption does not hold universally, and that the UV-absorber column density derived from the apparent optical depth of the observed lines should be treated rather as a lower limit. The real values could be as much as two orders of magnitude higher, which could fully explain the apparent incongruity in Mrk\,110 as well.

Another system of blueshifted absorption lines, which we also identified in the COS spectrum, does not have a detectable counterpart in the RGS data. At the blueshift corresponding to $-6793 ~\mathrm{km\,s^{-1}}$ in the Mrk\,110 rest frame, clear signatures of \ion{Ly}{$\alpha$}, the \ion{C}{iv} doublet, and \ion{Si}{iv} are present. Unfortunately, it is not possible to place reliable constraints on the physical properties of the gas producing these features. Finally, \ion{Ly}{$\alpha$} absorption lines at 1254.5, 1255.5, and 1258.0~\AA\  with no detected counterparts cannot be assigned to a well-characterised source either, which might be intrinsic to the Mrk\,110 environment or may originate in the intergalactic medium. 

\section{Summary}\label{sec:summary}
We analysed the X-ray, UV, and optical spectra of Mrk\,110 and performed global modelling of the BLR emission lines. The LOC model proved applicable to this NLS1 and the results suggest the following. 

\begin{enumerate}
\item The primarily UV-emitting gas can account also for the broad X-ray lines observed in this source. The only exception is \ion{Ne}{X}, which could belong to a different morphological component.
\item The BLR is more compact and less ionised than in the broad-line Seyfert 1s to which this approach has been applied so far. Furthermore, a distribution of the LOC clouds with a steeper radial profile is preferred by the data, reaching the dust sublimation radius. 
\item The intrinsic extinction associated with the emission lines is likely smaller than what the widely used Balmer ratio suggests. The LOC models for $E(B-V)_{\mathrm{lines}}\lesssim0.1$ are preferred at 3$\sigma$ significance, despite the Balmer ratio predicting as much as four times that value. The continuum-affecting extinction is more constrained ---thanks in part to the resulting SED shape--- to $0.01\lesssim E(B-V)_{\mathrm{SED}}\lesssim0.07$.
\item Finally, for a better description of the observed spectra, a model predicting a greater contribution from the brightest lines is necessary. This could be achieved if an equatorial clumpy wind component is added to the model.
\end{enumerate}

\begin{acknowledgements}
    The Space Research Organisation of the Netherlands is financially supported by NWO. Based on observations obtained with \textit{XMM-Newton}, a European Space Agency (ESA) science mission with instruments and contributions directly funded by ESA Member States and National Aeronautics Space Administration (NASA), and observations made with the NASA/ESA Hubble Space Telescope obtained from the Space Telescope Science Institute, which is operated by the Association of Universities for Research in Astronomy, Inc., under NASA contract NAS 5–26555. This research has made use of the NASA/IPAC Extragalactic Database (NED), which is operated by the Jet Propulsion Laboratory, California Institute of Technology, under contract with NASA. The data analysis and plots presented in this paper can be reproduced using the code and data files available at http://doi.org/10.5281/zenodo.10036304.
\end{acknowledgements}

\bibliographystyle{aa}
\bibliography{mrk110}

\begin{thebibliography}{78}
\expandafter\ifx\csname natexlab\endcsname\relax\def\natexlab#1{#1}\fi

\bibitem[{{Afanasiev} {et~al.}(2019){Afanasiev}, {Popovi{\'c}}, \&
  {Shapovalova}}]{Afanasiev2019}
{Afanasiev}, V.~L., {Popovi{\'c}}, L.~{\v{C}}., \& {Shapovalova}, A.~I. 2019,
  \mnras, 482, 4985

\bibitem[{{Akaike}(1974)}]{Akaike1974}
{Akaike}, H. 1974, IEEE Transactions on Automatic Control, 19, 716

\bibitem[{{Antonucci}(1993)}]{Antonucci1993}
{Antonucci}, R. 1993, \araa, 31, 473

\bibitem[{{Arav} {et~al.}(2002){Arav}, {Korista}, \& {de Kool}}]{Arav2002}
{Arav}, N., {Korista}, K.~T., \& {de Kool}, M. 2002, \apj, 566, 699

\bibitem[{{Baldwin} {et~al.}(1995){Baldwin}, {Ferland}, {Korista}, \&
  {Verner}}]{Baldwin1995}
{Baldwin}, J., {Ferland}, G., {Korista}, K., \& {Verner}, D. 1995, \apjl, 455,
  L119

\bibitem[{{Baldwin}(1975)}]{Baldwin1975}
{Baldwin}, J.~A. 1975, \apj, 201, 26

\bibitem[{{Baldwin}(1997)}]{Baldwin1997}
{Baldwin}, J.~A. 1997, in Astronomical Society of the Pacific Conference
  Series, Vol. 113, IAU Colloq. 159: Emission Lines in Active Galaxies: New
  Methods and Techniques, ed. B.~M. {Peterson}, F.-Z. {Cheng}, \& A.~S.
  {Wilson}, 80

\bibitem[{{Bianchi} {et~al.}(2005){Bianchi}, {Matt}, {Nicastro}, {Porquet}, \&
  {Dubau}}]{Bianchi2005}
{Bianchi}, S., {Matt}, G., {Nicastro}, F., {Porquet}, D., \& {Dubau}, J. 2005,
  \mnras, 357, 599

\bibitem[{{Bischoff} \& {Kollatschny}(1999)}]{Bischoff1999}
{Bischoff}, K. \& {Kollatschny}, W. 1999, \aap, 345, 49

\bibitem[{{Boller} {et~al.}(1996){Boller}, {Brandt}, \& {Fink}}]{Boller1996}
{Boller}, T., {Brandt}, W.~N., \& {Fink}, H. 1996, \aap, 305, 53

\bibitem[{{Bottorff} {et~al.}(2002){Bottorff}, {Baldwin}, {Ferland},
  {Ferguson}, \& {Korista}}]{Bottorff2002}
{Bottorff}, M.~C., {Baldwin}, J.~A., {Ferland}, G.~J., {Ferguson}, J.~W., \&
  {Korista}, K.~T. 2002, \apj, 581, 932

\bibitem[{{Burrows} {et~al.}(2005){Burrows}, {Hill}, {Nousek}, {Kennea},
  {Wells}, {Osborne}, {Abbey}, {Beardmore}, {Mukerjee}, {Short}, {Chincarini},
  {Campana}, {Citterio}, {Moretti}, {Pagani}, {Tagliaferri}, {Giommi},
  {Capalbi}, {Tamburelli}, {Angelini}, {Cusumano}, {Br{\"a}uninger}, {Burkert},
  \& {Hartner}}]{Burrows2005}
{Burrows}, D.~N., {Hill}, J.~E., {Nousek}, J.~A., {et~al.} 2005, \ssr, 120, 165

\bibitem[{{Cash}(1979)}]{Cash1979}
{Cash}, W. 1979, \apj, 228, 939

\bibitem[{{Costantini} {et~al.}(2007){Costantini}, {Kaastra}, {Arav}, {Kriss},
  {Steenbrugge}, {Gabel}, {Verbunt}, {Behar}, {Gaskell}, {Korista}, {Proga},
  {Quijano}, {Scott}, {Klimek}, \& {Hedrick}}]{Costantini2007}
{Costantini}, E., {Kaastra}, J.~S., {Arav}, N., {et~al.} 2007, \aap, 461, 121

\bibitem[{{Costantini} {et~al.}(2010){Costantini}, {Kaastra}, {Korista},
  {Ebrero}, {Arav}, {Kriss}, \& {Steenbrugge}}]{Costantini2010}
{Costantini}, E., {Kaastra}, J.~S., {Korista}, K., {et~al.} 2010, \aap, 512,
  A25

\bibitem[{{Costantini} {et~al.}(2016){Costantini}, {Kriss}, {Kaastra},
  {Bianchi}, {Branduardi-Raymont}, {Cappi}, {De Marco}, {Ebrero}, {Mehdipour},
  {Petrucci}, {Paltani}, {Ponti}, {Steenbrugge}, \& {Arav}}]{Costantini2016}
{Costantini}, E., {Kriss}, G., {Kaastra}, J.~S., {et~al.} 2016, \aap, 595, A106

\bibitem[{{Davidson}(1977)}]{Davidson1977}
{Davidson}, K. 1977, \apj, 218, 20

\bibitem[{{den Herder} {et~al.}(2001){den Herder}, {Brinkman}, {Kahn},
  {Branduardi-Raymont}, {Thomsen}, {Aarts}, {Audard}, {Bixler}, {den Boggende},
  {Cottam}, {Decker}, {Dubbeldam}, {Erd}, {Goulooze}, {G{\"u}del}, {Guttridge},
  {Hailey}, {Janabi}, {Kaastra}, {de Korte}, {van Leeuwen}, {Mauche},
  {McCalden}, {Mewe}, {Naber}, {Paerels}, {Peterson}, {Rasmussen}, {Rees},
  {Sakelliou}, {Sako}, {Spodek}, {Stern}, {Tamura}, {Tandy}, {de Vries},
  {Welch}, \& {Zehnder}}]{RGS}
{den Herder}, J.~W., {Brinkman}, A.~C., {Kahn}, S.~M., {et~al.} 2001, \aap,
  365, L7

\bibitem[{{Done} {et~al.}(2012){Done}, {Davis}, {Jin}, {Blaes}, \&
  {Ward}}]{Done2012}
{Done}, C., {Davis}, S.~W., {Jin}, C., {Blaes}, O., \& {Ward}, M. 2012, \mnras,
  420, 1848

\bibitem[{{Dong} {et~al.}(2008){Dong}, {Wang}, {Wang}, {Yuan}, {Zhou}, {Dai},
  \& {Zhang}}]{Dong2008}
{Dong}, X., {Wang}, T., {Wang}, J., {et~al.} 2008, \mnras, 383, 581

\bibitem[{{Ferguson} {et~al.}(1997){Ferguson}, {Korista}, {Baldwin}, \&
  {Ferland}}]{Ferguson1997}
{Ferguson}, J.~W., {Korista}, K.~T., {Baldwin}, J.~A., \& {Ferland}, G.~J.
  1997, \apj, 487, 122

\bibitem[{{Ferland} {et~al.}(2017){Ferland}, {Chatzikos}, {Guzm{\'a}n},
  {Lykins}, {van Hoof}, {Williams}, {Abel}, {Badnell}, {Keenan}, {Porter}, \&
  {Stancil}}]{Ferland2017}
{Ferland}, G.~J., {Chatzikos}, M., {Guzm{\'a}n}, F., {et~al.} 2017, \rmxaa, 53,
  385

\bibitem[{{Fitzpatrick}(1999)}]{Fitzpatrick1999}
{Fitzpatrick}, E.~L. 1999, \pasp, 111, 63

\bibitem[{{Gehrels} {et~al.}(2004){Gehrels}, {Chincarini}, {Giommi}, {Mason},
  {Nousek}, {Wells}, {White}, {Barthelmy}, {Burrows}, {Cominsky}, {Hurley},
  {Marshall}, {M{\'e}sz{\'a}ros}, {Roming}, {Angelini}, {Barbier}, {Belloni},
  {Campana}, {Caraveo}, {Chester}, {Citterio}, {Cline}, {Cropper}, {Cummings},
  {Dean}, {Feigelson}, {Fenimore}, {Frail}, {Fruchter}, {Garmire}, {Gendreau},
  {Ghisellini}, {Greiner}, {Hill}, {Hunsberger}, {Krimm}, {Kulkarni}, {Kumar},
  {Lebrun}, {Lloyd-Ronning}, {Markwardt}, {Mattson}, {Mushotzky}, {Norris},
  {Osborne}, {Paczynski}, {Palmer}, {Park}, {Parsons}, {Paul}, {Rees},
  {Reynolds}, {Rhoads}, {Sasseen}, {Schaefer}, {Short}, {Smale}, {Smith},
  {Stella}, {Tagliaferri}, {Takahashi}, {Tashiro}, {Townsley}, {Tueller},
  {Turner}, {Vietri}, {Voges}, {Ward}, {Willingale}, {Zerbi}, \&
  {Zhang}}]{Gehrels2004}
{Gehrels}, N., {Chincarini}, G., {Giommi}, P., {et~al.} 2004, \apj, 611, 1005

\bibitem[{{Goad} {et~al.}(2012){Goad}, {Korista}, \& {Ruff}}]{Goad2012}
{Goad}, M.~R., {Korista}, K.~T., \& {Ruff}, A.~J. 2012, \mnras, 426, 3086

\bibitem[{{Goodrich}(1989)}]{Goodrich1989}
{Goodrich}, R.~W. 1989, \apj, 342, 224

\bibitem[{{Gordon} {et~al.}(2003){Gordon}, {Clayton}, {Misselt}, {Landolt}, \&
  {Wolff}}]{Gordon2003}
{Gordon}, K.~D., {Clayton}, G.~C., {Misselt}, K.~A., {Landolt}, A.~U., \&
  {Wolff}, M.~J. 2003, \apj, 594, 279

\bibitem[{{Grupe} {et~al.}(2010){Grupe}, {Komossa}, {Leighly}, \&
  {Page}}]{Grupe2010}
{Grupe}, D., {Komossa}, S., {Leighly}, K.~M., \& {Page}, K.~L. 2010, \apjs,
  187, 64

\bibitem[{{Grupe} \& {Mathur}(2004)}]{Grupe2004a}
{Grupe}, D. \& {Mathur}, S. 2004, \apjl, 606, L41

\bibitem[{{Grupe} {et~al.}(2004){Grupe}, {Wills}, {Leighly}, \&
  {Meusinger}}]{Grupe2004}
{Grupe}, D., {Wills}, B.~J., {Leighly}, K.~M., \& {Meusinger}, H. 2004, \aj,
  127, 156

\bibitem[{{Homan} {et~al.}(2022){Homan}, {Lawrence}, {Ward}, {Bruce}, {Landt},
  {MacLeod}, {Elvis}, {Wilkes}, {Huchra}, \& {Peterson}}]{Homan2022}
{Homan}, D., {Lawrence}, A., {Ward}, M., {et~al.} 2022, \mnras
  [\eprint[arXiv]{2212.00684}]

\bibitem[{{Jaffarian} \& {Gaskell}(2020)}]{Jaffarian2020}
{Jaffarian}, G.~W. \& {Gaskell}, C.~M. 2020, \mnras, 493, 930

\bibitem[{{Jansen} {et~al.}(2001){Jansen}, {Lumb}, {Altieri}, {Clavel}, {Ehle},
  {Erd}, {Gabriel}, {Guainazzi}, {Gondoin}, {Much}, {Munoz}, {Santos},
  {Schartel}, {Texier}, \& {Vacanti}}]{XMM}
{Jansen}, F., {Lumb}, D., {Altieri}, B., {et~al.} 2001, \aap, 365, L1

\bibitem[{{Jha} {et~al.}(2022){Jha}, {Chand}, {Ojha}, {Omar}, \&
  {Rastogi}}]{Jha2022}
{Jha}, V.~K., {Chand}, H., {Ojha}, V., {Omar}, A., \& {Rastogi}, S. 2022,
  \mnras, 510, 4379

\bibitem[{{Kaastra}(2017)}]{Kaastra2017}
{Kaastra}, J.~S. 2017, \aap, 605, A51

\bibitem[{{Kaastra} {et~al.}(1996){Kaastra}, {Mewe}, \&
  {Nieuwenhuijzen}}]{SPEX}
{Kaastra}, J.~S., {Mewe}, R., \& {Nieuwenhuijzen}, H. 1996, in UV and X-ray
  Spectroscopy of Astrophysical and Laboratory Plasmas, ed. K.~{Yamashita} \&
  T.~{Watanabe} (Universal Academy Press, Tokyo), 411--414

\bibitem[{Kaastra {et~al.}(2022)Kaastra, Raassen, de~Plaa, \& Gu}]{SPEX30701}
Kaastra, J.~S., Raassen, A. J.~J., de~Plaa, J., \& Gu, L. 2022, SPEX X-ray
  spectral fitting package

\bibitem[{{Kalberla} {et~al.}(2005){Kalberla}, {Burton}, {Hartmann}, {Arnal},
  {Bajaja}, {Morras}, \& {P{\"o}ppel}}]{LAB}
{Kalberla}, P.~M.~W., {Burton}, W.~B., {Hartmann}, D., {et~al.} 2005, \aap,
  440, 775

\bibitem[{{Kallman} \& {Bautista}(2001)}]{Kallman2001}
{Kallman}, T. \& {Bautista}, M. 2001, \apjs, 133, 221

\bibitem[{{Keel}(1996)}]{Keel1996}
{Keel}, W.~C. 1996, \aj, 111, 696

\bibitem[{{Kinney} {et~al.}(1996){Kinney}, {Calzetti}, {Bohlin}, {McQuade},
  {Storchi-Bergmann}, \& {Schmitt}}]{Kinney1996}
{Kinney}, A.~L., {Calzetti}, D., {Bohlin}, R.~C., {et~al.} 1996, \apj, 467, 38

\bibitem[{{Kollatschny} {et~al.}(2001){Kollatschny}, {Bischoff}, {Robinson},
  {Welsh}, \& {Hill}}]{Kollatschny2001}
{Kollatschny}, W., {Bischoff}, K., {Robinson}, E.~L., {Welsh}, W.~F., \&
  {Hill}, G.~J. 2001, \aap, 379, 125

\bibitem[{{Korista} {et~al.}(1997){Korista}, {Baldwin}, {Ferland}, \&
  {Verner}}]{Korista1997a}
{Korista}, K., {Baldwin}, J., {Ferland}, G., \& {Verner}, D. 1997, \apjs, 108,
  401

\bibitem[{{Korista} \& {Goad}(2000)}]{Korista2000}
{Korista}, K.~T. \& {Goad}, M.~R. 2000, \apj, 536, 284

\bibitem[{{Korista} \& {Goad}(2004)}]{Korista2004}
{Korista}, K.~T. \& {Goad}, M.~R. 2004, \apj, 606, 749

\bibitem[{{Krause} {et~al.}(2012){Krause}, {Schartmann}, \&
  {Burkert}}]{Krause2012}
{Krause}, M., {Schartmann}, M., \& {Burkert}, A. 2012, \mnras, 425, 3172

\bibitem[{{Kubota} \& {Done}(2018)}]{Kubota2018}
{Kubota}, A. \& {Done}, C. 2018, \mnras, 480, 1247

\bibitem[{{Landt} {et~al.}(2014){Landt}, {Ward}, {Elvis}, \&
  {Karovska}}]{Landt2014}
{Landt}, H., {Ward}, M.~J., {Elvis}, M., \& {Karovska}, M. 2014, \mnras, 439,
  1051

\bibitem[{{Laor} \& {Draine}(1993)}]{Laor1993}
{Laor}, A. \& {Draine}, B.~T. 1993, \apj, 402, 441

\bibitem[{{Leighly} \& {Moore}(2004)}]{Leighly2004}
{Leighly}, K.~M. \& {Moore}, J.~R. 2004, \apj, 611, 107

\bibitem[{{Mason} {et~al.}(2001){Mason}, {Breeveld}, {Much}, {Carter},
  {Cordova}, {Cropper}, {Fordham}, {Huckle}, {Ho}, {Kawakami}, {Kennea},
  {Kennedy}, {Mittaz}, {Pandel}, {Priedhorsky}, {Sasseen}, {Shirey}, {Smith},
  \& {Vreux}}]{OM}
{Mason}, K.~O., {Breeveld}, A., {Much}, R., {et~al.} 2001, \aap, 365, L36

\bibitem[{{Matthews} {et~al.}(2020){Matthews}, {Knigge}, {Higginbottom},
  {Long}, {Sim}, {Mangham}, {Parkinson}, \& {Hewitt}}]{Matthews2020}
{Matthews}, J.~H., {Knigge}, C., {Higginbottom}, N., {et~al.} 2020, \mnras,
  492, 5540

\bibitem[{{Mor} \& {Netzer}(2012)}]{MorNetzer2012}
{Mor}, R. \& {Netzer}, H. 2012, \mnras, 420, 526

\bibitem[{{Nagao} {et~al.}(2006){Nagao}, {Marconi}, \& {Maiolino}}]{Nagao2006}
{Nagao}, T., {Marconi}, A., \& {Maiolino}, R. 2006, \aap, 447, 157

\bibitem[{{Netzer} \& {Laor}(1993)}]{Netzer1993}
{Netzer}, H. \& {Laor}, A. 1993, \apjl, 404, L51

\bibitem[{{Osterbrock}(1989)}]{Osterbrock1989}
{Osterbrock}, D.~E. 1989, {Astrophysics of gaseous nebulae and active galactic
  nuclei} (University Science Books)

\bibitem[{{Pancoast} {et~al.}(2012){Pancoast}, {Brewer}, {Treu}, {Barth},
  {Bennert}, {Canalizo}, {Filippenko}, {Gates}, {Greene}, {Li}, {Malkan},
  {Sand}, {Stern}, {Woo}, {Assef}, {Bae}, {Buehler}, {Cenko}, {Clubb},
  {Cooper}, {Diamond-Stanic}, {Hiner}, {H{\"o}nig}, {Joner}, {Kandrashoff},
  {Laney}, {Lazarova}, {Nierenberg}, {Park}, {Silverman}, {Son}, {Sonnenfeld},
  {Thorman}, {Tollerud}, {Walsh}, \& {Walters}}]{Pancoast2012}
{Pancoast}, A., {Brewer}, B.~J., {Treu}, T., {et~al.} 2012, \apj, 754, 49

\bibitem[{{Pancoast} {et~al.}(2014){Pancoast}, {Brewer}, {Treu}, {Park},
  {Barth}, {Bentz}, \& {Woo}}]{Pancoast2014}
{Pancoast}, A., {Brewer}, B.~J., {Treu}, T., {et~al.} 2014, \mnras, 445, 3073

\bibitem[{{Peretz} {et~al.}(2019){Peretz}, {Miller}, \& {Behar}}]{Peretz2019}
{Peretz}, U., {Miller}, J.~M., \& {Behar}, E. 2019, \apj, 879, 102

\bibitem[{{Perola} {et~al.}(2002){Perola}, {Matt}, {Cappi}, {Fiore},
  {Guainazzi}, {Maraschi}, {Petrucci}, \& {Piro}}]{Perola2002}
{Perola}, G.~C., {Matt}, G., {Cappi}, M., {et~al.} 2002, \aap, 389, 802

\bibitem[{{Peterson}(1993)}]{Peterson1993}
{Peterson}, B.~M. 1993, \pasp, 105, 247

\bibitem[{{Peterson} {et~al.}(1998){Peterson}, {Wanders}, {Bertram}, {Hunley},
  {Pogge}, \& {Wagner}}]{Peterson1998}
{Peterson}, B.~M., {Wanders}, I., {Bertram}, R., {et~al.} 1998, \apj, 501, 82

\bibitem[{{Ponti} {et~al.}(2013){Ponti}, {Cappi}, {Costantini}, {Bianchi},
  {Kaastra}, {De Marco}, {Fender}, {Petrucci}, {Kriss}, {Steenbrugge}, {Arav},
  {Behar}, {Branduardi-Raymont}, {Dadina}, {Ebrero}, {Lubi{\'n}ski},
  {Mehdipour}, {Paltani}, {Pinto}, \& {Tombesi}}]{Ponti2013}
{Ponti}, G., {Cappi}, M., {Costantini}, E., {et~al.} 2013, \aap, 549, A72

\bibitem[{{Porquet} \& {Dubau}(2000)}]{Porquet2000}
{Porquet}, D. \& {Dubau}, J. 2000, \aaps, 143, 495

\bibitem[{{Reeves} {et~al.}(2021){Reeves}, {Porquet}, {Braito}, {Grosso}, \&
  {Lobban}}]{Reeves2021}
{Reeves}, J.~N., {Porquet}, D., {Braito}, V., {Grosso}, N., \& {Lobban}, A.
  2021, \aap, 649, L3

\bibitem[{{Roming} {et~al.}(2005){Roming}, {Kennedy}, {Mason}, {Nousek}, {Ahr},
  {Bingham}, {Broos}, {Carter}, {Hancock}, {Huckle}, {Hunsberger}, {Kawakami},
  {Killough}, {Koch}, {McLelland}, {Smith}, {Smith}, {Soto}, {Boyd},
  {Breeveld}, {Holland}, {Ivanushkina}, {Pryzby}, {Still}, \&
  {Stock}}]{Roming2005}
{Roming}, P. W.~A., {Kennedy}, T.~E., {Mason}, K.~O., {et~al.} 2005, \ssr, 120,
  95

\bibitem[{{Schlafly} \& {Finkbeiner}(2011)}]{Schlafly2011}
{Schlafly}, E.~F. \& {Finkbeiner}, D.~P. 2011, \apj, 737, 103

\bibitem[{{Suganuma} {et~al.}(2006){Suganuma}, {Yoshii}, {Kobayashi},
  {Minezaki}, {Enya}, {Tomita}, {Aoki}, {Koshida}, \&
  {Peterson}}]{Suganuma2006}
{Suganuma}, M., {Yoshii}, Y., {Kobayashi}, Y., {et~al.} 2006, \apj, 639, 46

\bibitem[{{Sulentic} {et~al.}(2000){Sulentic}, {Zwitter}, {Marziani}, \&
  {Dultzin-Hacyan}}]{Sulentic2000}
{Sulentic}, J.~W., {Zwitter}, T., {Marziani}, P., \& {Dultzin-Hacyan}, D. 2000,
  \apjl, 536, L5

\bibitem[{{Tsuzuki} {et~al.}(2006){Tsuzuki}, {Kawara}, {Yoshii}, {Oyabu},
  {Tanab{\'e}}, \& {Matsuoka}}]{Tsuzuki2006}
{Tsuzuki}, Y., {Kawara}, K., {Yoshii}, Y., {et~al.} 2006, \apj, 650, 57

\bibitem[{{U} {et~al.}(2022){U}, {Barth}, {Vogler}, {Guo}, {Treu}, {Bennert},
  {Canalizo}, {Filippenko}, {Gates}, {Hamann}, {Joner}, {Malkan}, {Pancoast},
  {Williams}, {Woo}, {Abolfathi}, {Abramson}, {Armen}, {Bae}, {Bohn},
  {Boizelle}, {Bostroem}, {Brandel}, {Brink}, {Channa}, {Cooper}, {Cosens},
  {Donohue}, {Fillingham}, {Gonz{\'a}lez-Buitrago}, {Halevi}, {Halle}, {Hood},
  {Horne}, {Horst}, {Kouchkovsky}, {Kuhn}, {Kumar}, {Leonard}, {Loveland},
  {Manzano-King}, {McHardy}, {Michel}, {Olaes}, {Park}, {Park}, {Pei}, {Ross},
  {Runco}, {Samuel}, {S{\'a}nchez}, {Scott}, {Sexton}, {Shin}, {Shivvers},
  {Spencer}, {Stahl}, {Stegman}, {Stomberg}, {Valenti}, {Villafa{\~n}a},
  {Walsh}, {Yuk}, \& {Zheng}}]{U2022}
{U}, V., {Barth}, A.~J., {Vogler}, H.~A., {et~al.} 2022, \apj, 925, 52

\bibitem[{{Vasudevan} \& {Fabian}(2009)}]{Vasudevan2009}
{Vasudevan}, R.~V. \& {Fabian}, A.~C. 2009, \mnras, 392, 1124

\bibitem[{{Villafa{\~n}a} {et~al.}(2022){Villafa{\~n}a}, {Williams}, {Treu},
  {Brewer}, {Barth}, {U}, {Bennert}, {Vogler}, {Guo}, {Bentz}, {Canalizo},
  {Filippenko}, {Gates}, {Hamann}, {Joner}, {Malkan}, {Woo}, {Abolfathi},
  {Abramson}, {Armen}, {Bae}, {Bohn}, {Boizelle}, {Bostroem}, {Brandel},
  {Brink}, {Channa}, {Cooper}, {Cosens}, {Donohue}, {Fillingham},
  {Gonzalez-Buitrago}, {Halevi}, {Halle}, {Hood}, {Horne}, {Horst}, {de
  Kouchkovsky}, {Kuhn}, {Kumar}, {Leonard}, {Loveland}, {Manzano-King},
  {McHardy}, {Michel}, {Olaes}, {Park}, {Park}, {Pei}, {Ross}, {Runco},
  {Samuel}, {Sanchez}, {Scott}, {Sexton}, {Shin}, {Shivvers}, {Spencer},
  {Stahl}, {Stegman}, {Stomberg}, {Valenti}, {Walsh}, {Yuk}, \&
  {Zheng}}]{Villafana2022}
{Villafa{\~n}a}, L., {Williams}, P.~R., {Treu}, T., {et~al.} 2022, arXiv
  e-prints, arXiv:2203.15000

\bibitem[{{Vincentelli} {et~al.}(2021){Vincentelli}, {McHardy}, {Cackett},
  {Barth}, {Horne}, {Goad}, {Korista}, {Gelbord}, {Brandt}, {Edelson},
  {Miller}, {Pahari}, {Peterson}, {Schmidt}, {Baldi}, {Breedt}, {Hern{\'a}ndez
  Santisteban}, {Romero-Colmenero}, {Ward}, \& {Williams}}]{Vincentelli2021}
{Vincentelli}, F.~M., {McHardy}, I., {Cackett}, E.~M., {et~al.} 2021, \mnras,
  504, 4337

\bibitem[{{Wandel} {et~al.}(1999){Wandel}, {Peterson}, \&
  {Malkan}}]{Wandel1999}
{Wandel}, A., {Peterson}, B.~M., \& {Malkan}, M.~A. 1999, \apj, 526, 579

\bibitem[{{Waters} {et~al.}(2021){Waters}, {Proga}, \& {Dannen}}]{Waters2021}
{Waters}, T., {Proga}, D., \& {Dannen}, R. 2021, \apj, 914, 62

\bibitem[{{Winter} {et~al.}(2010){Winter}, {Lewis}, {Koss}, {Veilleux},
  {Keeney}, \& {Mushotzky}}]{Winter2010}
{Winter}, L.~M., {Lewis}, K.~T., {Koss}, M., {et~al.} 2010, \apj, 710, 503

\bibitem[{{Wu} \& {Han}(2001)}]{Wu2001}
{Wu}, X.-B. \& {Han}, J.~L. 2001, \apjl, 561, L59

\end{thebibliography}

\begin{appendix}
\section{X-ray line equivalent widths}

\begin{figure*}
\centering
\includegraphics[width=\textwidth]{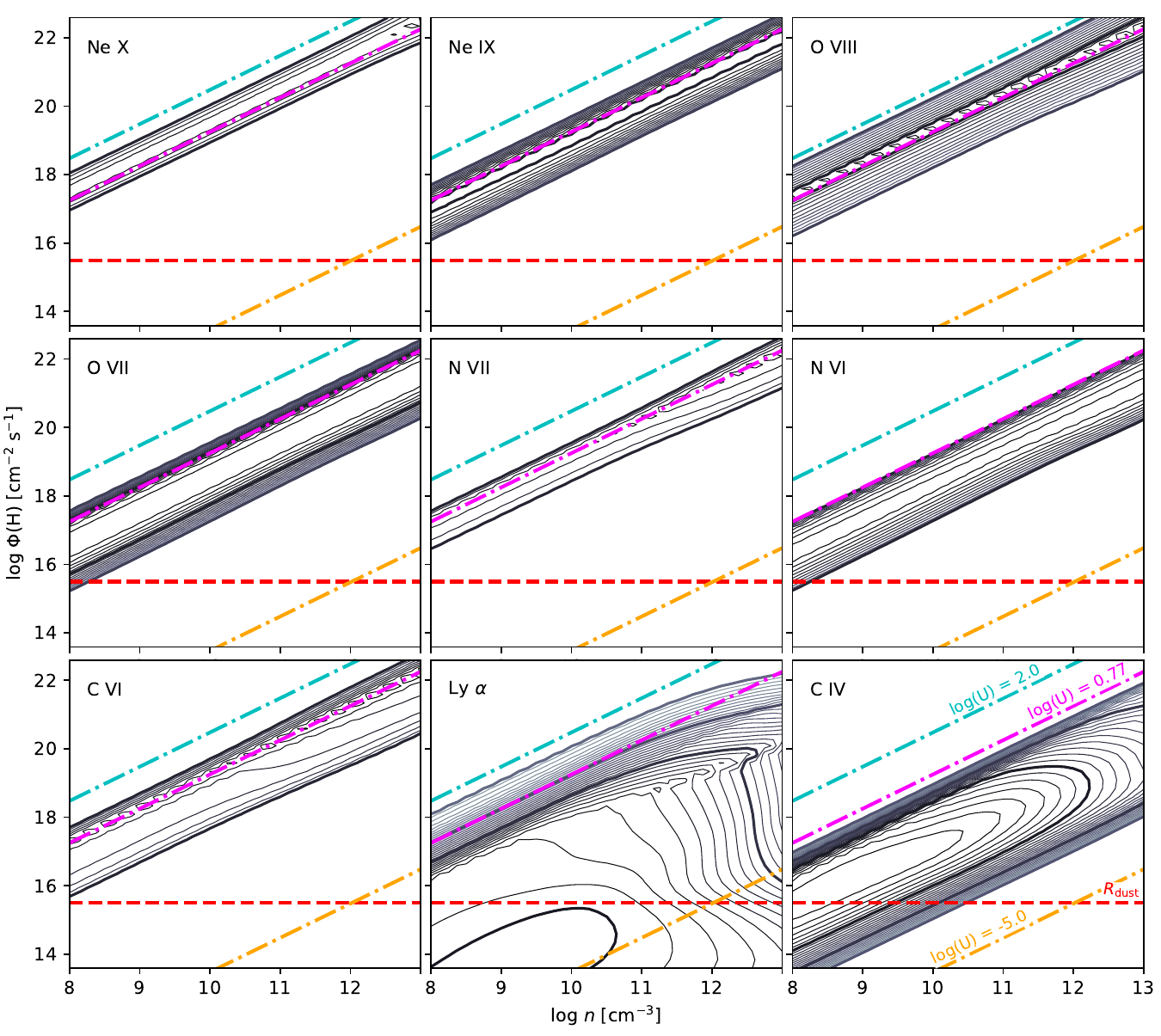}
\caption{Predicted line emission of the X-ray lines well-detected in Mrk\,110, \ion{C}{iv}, and \ion{Ly}{$\alpha$}, plotted as a function of the hydrogen-ionising photon flux and the gas density. The contours present the logarithmic line equivalent widths, referenced to the incident continuum at 1216~\AA\ and unity covering fraction. The contours plotted as thick solid lines correspond to 1~dex steps, starting with the smallest visualised value of 1~\AA. The thin lines, separated with steps of 0.1~dex, provide further details of the inwards increase in the emission within the typically rather narrow portions of the parameter space. The coloured lines are defined in the bottom-right box and are explained in Sect. \ref{ssec:locdefinition}.}
\label{fig:W-phi-n}
\end{figure*}

In Fig. \ref{fig:W-phi-n}, we show the predictions for line emission for the baseline model grid (see Sect. \ref{ssec:locdefinition}), displayed as a function of hydrogen particle density of the gas and the ionising photon flux. The contour plots are presented for the well-detected X-ray lines in the Mrk\,110 RGS spectrum and the two brightest UV lines, \ion{Ly}{$\alpha$} and \ion{C}{iv}, for comparison. For a detailed discussion of the predictions for the UV and optical lines, we refer the reader to \citet{Korista1997a}, where the aspects of the modelled emission are examined for a range of physical conditions. The plot also shows the limits (introduced in Sect. \ref{ssec:locdefinition}) placed on the cloud minimal and maximal ionisation and the maximal outer radius of the broad line region, which is applied to the line-luminosity fitting.\\

\section{Reconstruction of the Mrk\,110 SED vs. intrinsic extinction}
\begin{figure*}
\centering
\resizebox{\hsize}{!}{\includegraphics{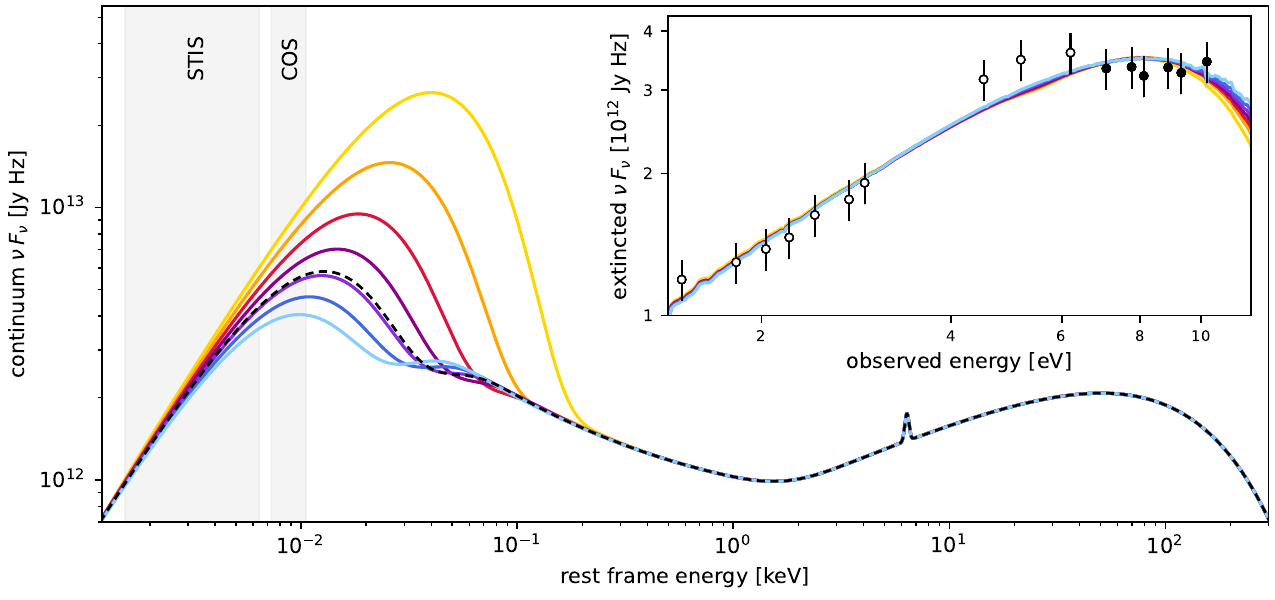}}
\caption{Extinction-dependent estimates of the Mrk\,110 SED. \textit{Inset:} UV and optical continuum points extracted from the HST COS (full symbols) and STIS (empty symbols) Galactic extinction-corrected spectra, fitted with the SED model discussed in Sect. \ref{ssec:extinction}. Each of the solid lines represents a model with different $E(B-V)$ of extinction intrinsic to the host object, ranging from 0.01 (light blue) to 0.07 (yellow). \textit{Main panel:} BLR-ionising radiation SEDs resulting from these fits, varying by the temperature and normalisation of the disc black body component. Energy bands covered by the COS and STIS data points are highlighted. The black dashed line represents the baseline SED with the disc black body temperature of $kT = 10~\rm{eV}$ and $E(B-V) = 0.032 \pm 0.001$.}
\label{fig:extSED}
\end{figure*}

In Fig. \ref{fig:extSED}, we give the SED shapes recovered using different amounts of extinction corrections applied to the continuum data points from the STIS and COS observations. In none of the models does the disc black-body contribute noticeably to the X-ray soft excess, which is dominated by the Comptonisation component with a fixed normalisation and the electron temperature. Assuming the black-hole mass of $2\times 10^7 \, M_{\odot}$, the Eddington ratio $L_{\mathrm{bol}}/L_{\mathrm{Edd}}$ corresponding to the displayed SEDs ranges from 0.28 for $E(B-V) = 0.01$ to 0.77 for the most luminous case, with $E(B-V) = 0.07$. For the baseline SED, $L_{\mathrm{bol}}/L_{\mathrm{Edd}} \approx 0.31$.\\

\end{appendix}

\end{document}